\documentclass[fleqn,usenatbib]{mn2e}

\usepackage{longtable}
\usepackage{graphicx}
\usepackage{layouts}
\usepackage{caption}
\usepackage{amsmath}
\usepackage{pdflscape}
\usepackage{color}
\usepackage{textgreek}
\usepackage[title,titletoc]{appendix}
\usepackage{mathtools}
\usepackage{footmisc}
\usepackage{subcaption}
\usepackage[md]{titlesec}
\usepackage{csvsimple}
\usepackage{epstopdf} 
\usepackage{epsfig}
\usepackage{pbox}

\DeclareGraphicsExtensions{.pdf,.png,.jpg,.eps,.ps}
 
\usepackage{graphicx}
\usepackage{float}

\def\aj{AJ}                   % Astronomical Journal
\def\araa{ARA\&A}             % Annual Review of Astron and Astrophys
\def\apj{ApJ}                 % Astrophysical Journal
\def\apjs{ApJS}               % Astrophysical Journal, Supplement
\def\aap{A\&A}                % Astronomy and Astrophysics
\def\mnras{MNRAS}             % Monthly Notices of the RAS
\def\pasp{PASP}               % Publications of the ASP
\def\pasj{PASJ}               % Publications of the ASJ
\def\vizier{VizieR Online Data Catalog}  %VizieR Online Data Catalog
\def\aat{Astronomical and Astrophysical Transactions}  %Astronomical and Astrophysical Transactions
\def\an{Astronomische Nachrichten}  %Astronomische Nachrichten
\def\arX{ArXiv e-prints}
\def\gaia{A\&A special Gaia volume}

\newcommand{\Msolar}{\mbox{\,$\rm M_{\odot}$}}        % solar mass
\newcommand{\kmsec}{\,km s$^{-1}$}
\newcommand{\lsfour}{LS\,IV$-14^\circ116$}

\title[Kinematics of sdO/B stars]{Kinematics of Subluminous O and B Stars by Surface Helium Abundance}

\author[P. Martin et al.]{P. Martin$^{1,2}$\thanks{E-mail: pam@arm.ac.uk}, C. S. Jeffery$^{1,2}$\thanks{E-mail: csj@arm.ac.uk}, Naslim N.$^{1,3}$ and V. M. Woolf$^{ 4}$ \\
$^{1}$Armagh Observatory and Planetarium, College Hill, Armagh, BT61 9DG, UK\\
$^{2}$School of Physics, Trinity College Dublin, College Green, Dublin 2, Ireland\\
$^{3}$Academia Sinica Institute of Astronomy and Astrophysics, Taipei 10617, Taiwan, Republic of China \\
$^{4}$Department of Physics, University of Nebraska at Omaha, 6001 Dodge Street, Omaha, NE 68182, USA}

\date{Accepted . Received ; in original form }

\begin{document}
\maketitle

\label{firstpage}

\begin{abstract}
The majority of hot subdwarf stars are low-mass core-helium-burning stars.
Their atmospheres are generally helium deficient; however a minority have extremely helium-rich surfaces. 
An additional fraction have an intermediate surface-helium abundance, occasionally accompanied by peculiar abundances of other elements.
We have identified a sample of { 88 hot subdwarfs including 38 helium-deficient, 27 intermediate-helium and 23 extreme-helium stars} for which radial-velocity and proper-motion measurements, together with distances, allow a calculation of Galactic space velocities. 
We have investigated the kinematics of these three groups to determine whether they belong to similar or different galactic populations. 
The majority of helium-deficient subdwarfs in our sample show a kinematic distribution  similar to that of thick disk stars. Helium-rich sdBs show a more diverse kinematic distribution. 
Although the majority are probably disk stars, a minority show a much higher velocity dispersion consistent with membership of a Galactic halo population. 
Several of the halo subdwarfs are members of the class of ``heavy-metal'' subdwarfs discovered by \citet{naslim11,naslim13}. 
\end{abstract}


\section{Introduction}

Hot subluminous stars or subdwarfs are traditionally classified into three types by their spectra \citep{drilling03}. Subdwarf B (sdB) stars have a surface effective temperature $T_{\rm eff}$ in the range 20,000--40,000 K and hydrogen-Balmer absorption lines wider than in normal B stars; subdwarf O (sdO) stars, with $T_{\rm eff}$ ranging from 40,000--80,000 K have strong He$^{\rm +}$ absorption lines; subdwarf OB (sdOB) stars represent a transition between O and B types \citep{moehler90,heber09}. These objects are located below the upper main sequence on a Hertzsprung--Russell (HR) diagram. Identified as low-mass core-helium burning stars with low-mass envelopes, they are also known as extreme horizontal branch (EHB) stars.

The atmospheres of sdB stars are generally helium deficient, as radiative levitation and gravitational settling cause helium to sink below the hydrogen-rich surface \citep{heber86}, deplete other light elements and enhance many heavy elements in the photosphere \citep*{otoole06}.
However almost 10\% of the total subdwarf population comprises stars with helium-rich atmospheres. The helium-rich subdwarfs may be further divided into extremely helium-rich stars   
and a small number of intermediate helium-rich stars, a number of which show extreme chemical peculiarities \citep{naslim11,naslim13}.   {A recent review of hot subluminous stars has been given by \citet{heber16}.}

The formation of hot subdwarfs offers several puzzles; they are observed as single stars, and as both close and wide binaries. The helium-deficient sdOs are likely the progeny of sdB stars in a post-EHB phase as they have diffusion-dominated abundance patterns and similar binary frequencies \citep{heber09}.  Binary interaction  via one or two common-envelope ejection phases provides a very promising mechanism to explain the many close binaries found amongst the sdB stars \citep{han1,han2}. Extreme-helium subdwarfs appear to be well explained by the merger of two helium white dwarfs \citep{zhang12}. However, it is harder to understand the intermediate-helium subdwarfs; few have been analysed and those that have are diverse \citep{naslim11,naslim12,naslim13}. 

As the atmospheric abundance patterns of hot subdwarfs are governed by diffusion processes, they cannot be used to establish population membership. However, stars in the Milky Way formed at different epochs belong to different populations, which can be distinguished via kinematic criteria.
Kinematical data gives us access to the motions of the different populations of stars in the Galaxy. Some components of the Milky Way are rapidly rotating with little dispersion in the velocities of the members while others show only little rotation but high dispersions.
Orbital eccentricity may also distinguish between younger and older stars as more gravitational interactions lead to larger deviations from originally circular orbits (if that is where the progenitors formed).

Members of the thin disk population are found close to the Galactic plane in low eccentricity orbits. Heliocentric velocities of thin disk stars in the solar neighbourhood are small. Stars of the thick disk population orbit around the Galactic centre on more eccentric orbits and are found at higher distances from the plane. Typically, velocities are larger, relative to the Local Standard of Rest, than for thin disk objects. Stars of the Galactic halo population (also known as population II stars) are often found at large distances from the galactic disc and their orbits are often highly eccentric. Halo stars as a population do not (or not much) participate in the galactic rotation. These stars include those with the highest heliocentric velocities.

Previous studies involving the kinematics of sdBs found that the majority are members of the disk, but that a minority are halo members \citep{altmann04}. \cite{randall15} have recently shown that the intermediate-helium sdB star \lsfour\ has halo kinematics. This raises the question of whether the helium-rich hot subdwarfs all belong 
to this halo minority and the "normal" sdO/Bs are a disk population. In this paper we collate radial velocities and proper motions for a significant sample of helium-deficient, intermediate-helium and extreme-helium subdwarfs, including a number of new measurements (\S\,2). From these we compute space velocities (\S\,3) and galactic orbits 
in order to identify the parent populations (\S\,4). We discuss some of the implications of these results for interpreting the origin of, in particular, the chemically-peculiar 
intermediate-helium sdBs  (\S\,5).

\begin{table*}
\center
\caption{Heliocentric radial velocities for 32 helium-rich hot subdwarfs.}
\begin{tabular}{|c|c|c|cccc}
\hline
Star&Instrument &Date&RV (\kmsec)&$\gamma$&$\pm$\\
\hline
HE\,0001--2443 	&UVES	&15/10/02 		&4.68	&3.98&0.71\\
                           		&&18/06/03		&3.27&&\\
HE\,0111--1526	&UVES	&18/12/01 		&6.56	& --21.83&28.39\\
                           		&&29/12/01		&--50.22&&\\
HE\,1135--1134	&UVES	&28/06/01		&	22.22&24.67&2.45\\
                           		&&24/07/01		&27.11&&\\
HE\,1136--2504	&UVES	&22/04/00 		&68.6	&59.39&9.22\\
                          		 &&17/05/00		&50.17&&\\
HE\,1238--1745	&UVES	&23/06/01 		&--10.58&	--7.87&2.72\\
                           		&&23/07/01		&--5.15&&\\
HE\,1256--2738	&UVES	&22/04/00 		&146.26&	140.46&5.8\\
                           		&&19/05/00		&134.66&&\\
HE\,1258+0113	&UVES	&19/05/00		&--62.29&	--42.69&19.6\\
                           		&&22/05/00		&--23.09&&\\
HE\,1310--2733	&UVES	&22/04/00 		&39.72&	41.54&1.82\\
                           		&&24/04/00		&43.36&&\\
HE\,2218--2026	&UVES	&	24/09/02		&--278.91&--278.86&1.45\\
                           		&&25/09/02		&--281.81&&\\
HE\,2359--2844	&UVES	&15/09/02		&--93.74&--90.71&3.03\\
                           		&&25/09/02		&--87.68&&\\                           
$\rm UVO\,0825+15$	&HDS	&03/06/15 		&56.4&&0.5\\
J092440.11+305013.16 &HDS	&03/06/15 	&2.7&&0.5\\
J160131.30+044027.00 &HDS	&03/06/15 	&--27.4&&0.9\\
J175137.44+371952.37 &HDS	&03/06/15 	&--73.6&&0.2\\
 J175548.50+501210.77 &HDS  &03/06/15  	&--62.7      &&0.2\\  
PG\,2321+214 		&HDS	&04/10/98 	&    --19.7 &       & 3 \\     

PG\,0902+057 &UES&06/05/95&--15&&5\\
PG\,1615+413&ISIS&31/05/96&--80&&5\\
PG\,1600+171&ISIS&31/05/96&--78&&5\\
PG\,1658+273&ISIS&31/05/96&--33&&5\\
PG\,1715+273&ISIS&31/05/96&31&&5\\
HS\,1844+637&ISIS&31/05/96&20&&10\\
PG\,1554+408&ISIS&31/05/96&73&&8\\
PG\,2258+155&ISIS&04/10/98&34&&5\\
PG\,1127+019&IDS&28/04/02&19&&2\\
PG\,1415+492&IDS&29/04/02&54&&1\\
PG\,2215+151&ISIS&31/05/96&--13&&5\\
HS\,1000+471  &ISIS   &31/05/96&0&&10\\

BPS\, CS 22956--0094&UCLES&27/08/05 &--4.1&&1\\
BPS\, CS 29496--0010 & UCLES&27/08/05 &  --39.8  && 0.1             \\
 BPS\, CS 22940--0009  & UCLES &26/08/05&47.8  && 0.5\\
 LB\, 3229        &UCLES&    27/08/05       & 42.7 & & 1.0  \\                          
\hline
\label{rv}
\end{tabular}
\label{distance}
\end{table*}

\section{Data}

We analysed a sample of 88 hot subdwarfs (sdO/B) including 38 helium deficient, 27 intermediate helium and 23 extreme helium stars. 
The criterion for inclusion was that each star should have measurements of radial velocity, helium abundance  and proper motion; 
these are shown in appendix Tables \ref{A1} -- \ref{A3}.

\subsection{\textit{Radial Velocities}}

Radial velocities for 63 stars were obtained from the literature, as described in Appendix Tables \ref{A1}--\ref{A3}. A small number 
of these have errors in excess of 30\kmsec; although large, the data were retained so as not to over-restrict the sample. The errors
were propagated through the space motion calculations.  

{Radial velocities for 32 helium-rich hot subdwarfs were measured by us and are presented here (Table \ref{rv}).}
Two spectra for 10 of these stars were obtained using the UVES instrument on the ESO VLT, as part of the ESO supernova progenitor survey between 2000 and 2003 
\citep{napiwotzki03}. Pipeline reduced spectra were recovered from the ESO archive. Velocity shifts were measured by cross-correlation with an appropriately chosen model spectrum. {Radial velocities for four stars observed in 2005 with the University College London Echelle Spectrograph (UCLES) on the Anglo-Australian Telescope (AAT),
 were measured by cross-correlation with theoretical spectra corresponding approximately to the solutions obtained by Naslim et al. (2010).}
The dates and heliocentric radial velocities measured for each observation are shown in Table \ref{rv}. 
with a mean uncertainty of $\pm$4.6\kmsec. 
The mean velocity for each star is represented as $\gamma$, and the error represents half the difference between the two 
measurements. For HE\,0111--1526 and  HE\,1258+0113,  these differences are, respectively, more than 8 and 5 times the mean differences of the remainder,
suggesting that one or both may be binaries.  {These stars have not been used in the kinematical analysis as the system velocity is unknown. }

{Five helium-rich sdO/B stars were observed with the HDS instrument on the {\it Subaru} 
telescope, operating in service mode on 2015 June 3. The data were reduced using 
standard IRAF procedures, the echelle orders were merged and wavelengths corrected for earth motion. 
Radial velocities of each star were measured by cross-correlation against a standard template. 
Two templates were used, being theoretical spectra for intermediate helium-rich hot subdwarfs 
and having effective  temperatures of 34\,000 and 40\,000 K, the first being carbon-rich, the second nitrogen-rich.  
Two spectral windows were used with each template, covering a part of both CCDs in the HDS instrument, 
and hence giving  four independent measurements of velocity for each observation. The means and standard 
deviations are given in Table 1. The radial velocities for the other twelve stars reported in Table 1 were measured 
from the blue part of the spectrum (4165 -- 4570 \AA) published by \cite{Ahmad03}, using cross-correlation 
as for the {\it Subaru} data. The template was an extremely helium-rich  model 
with effective temperature  40\,000\,K and surface gravity $\log g = 5.4$. }

Since binary companions affect the radial velocity of a star periodically, the system velocities, $\gamma$, were used for confirmed binary systems. 
We cannot exclude the possibility of unconfirmed binary systems which would affect the calculated Galactic velocities and orbits. 
Estimates of the binary fraction for EHB stars are between 50 and 60$\%$ (\citealt{maxted01}; \citealt{copper11}).
 
This is a lower bound since the radial velocity variations of very long period systems are difficult to detect.
{All stars with single epoch radial velocity measurements and confirmed binary systems are flagged in Tables A.9 -- A.11. These flagged stars should be considered as candidates for the thin disk, thick disk or halo populations until the radial velocities are confirmed.}

\begin{figure*}
  \begin{minipage}{180mm}
\begin{center}

\epsfig{file = 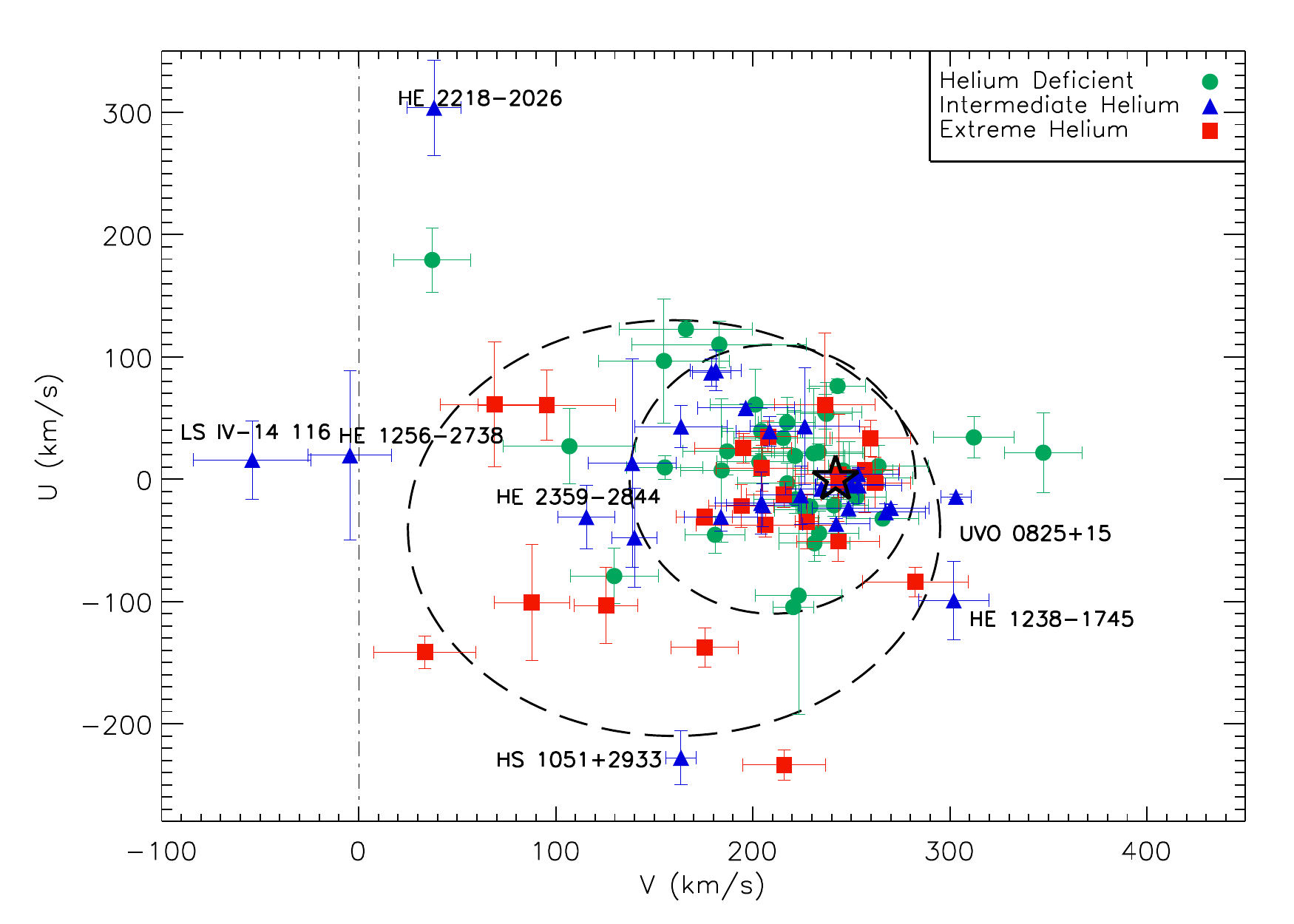, width=12cm}
\vspace{-0.5cm}
\caption{{$U-V$- velocity diagram with 3-$\sigma $ (thin disk) and 3-$\sigma$ (thick disk) contours. Red squares are extremely helium rich subdwarfs, blue triangles are intermediate helium rich subdwarfs and the green data circles are the helium deficient stars. The black star represents the Local Standard of Rest (LSR). The dot-dash line at a Galactic rotational velocity of zero is to highlight stars with retrograde motion.}}
\label{UV}
\end{center}
\end{minipage}
\end{figure*}

\begin{figure*}
\center
\includegraphics[width = 12cm]{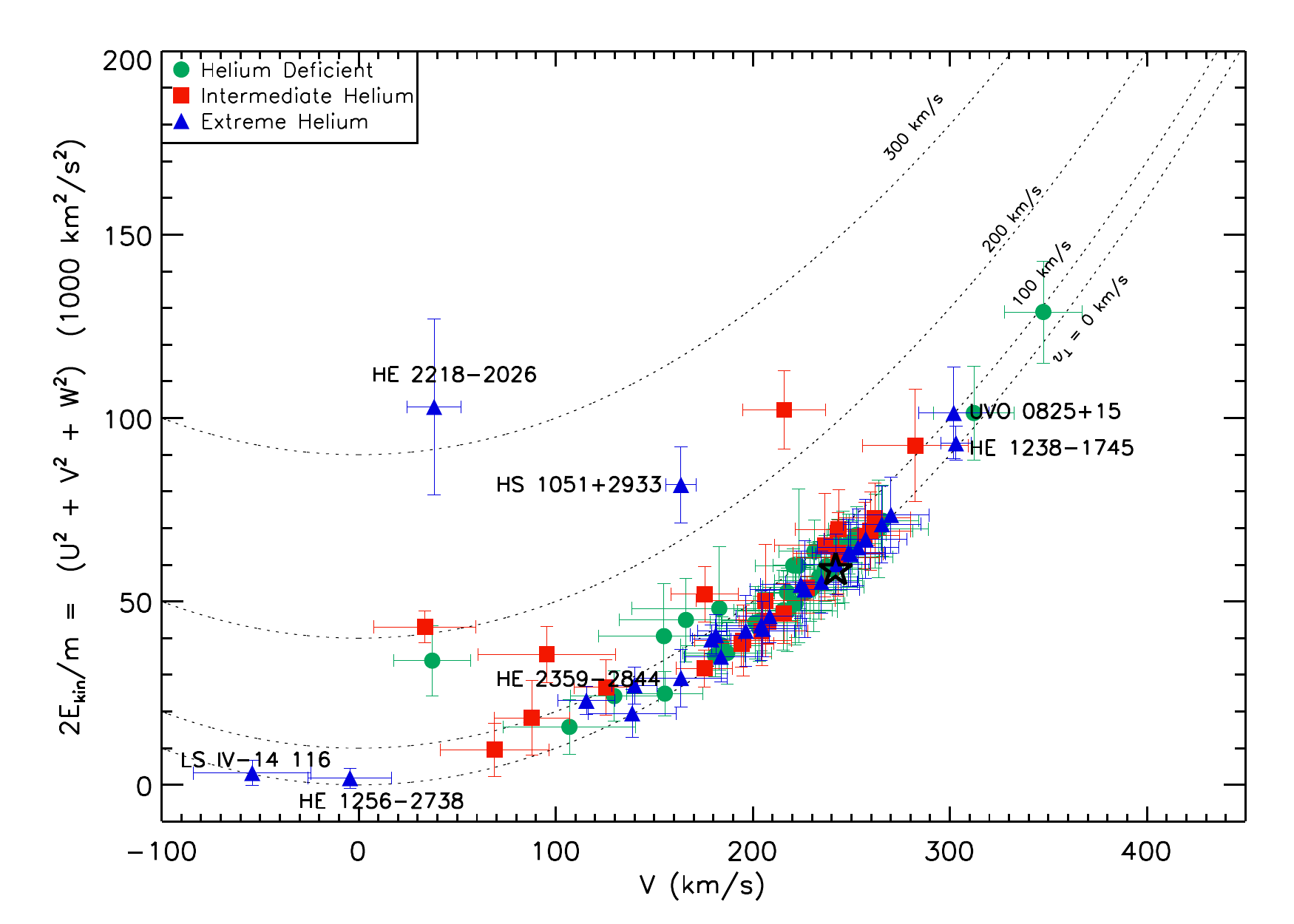}
\caption{{Galactic rotational velocity against the total kinetic energy. Symbols have the same meaning as above. The parabolic curves denote line of equal velocity $v_{\bot} = (U^2 + W^2)^{1/2}$. }}
\label{kinetic}
\end{figure*}

\subsection{\textit{Abundances}}

Subdwarf O and B stars can be further classified by their surface helium content. 
Based on the overall classification for hot subdwarfs developed by \citet{drilling13}, extreme helium-rich subdwarfs are defined to have $n_{\rm He} > 80\%$ by number, 
intermediate helium-rich subdwarfs have a helium abundance of $5\% < n_{\rm He} < 80\%$ 
and helium-deficient subdwarfs have $n_{\rm He} < 5\%$. 
Helium abundances for the current sample were obtained from \citet{ahmad07, edelmannphd, drilling13, muchfuss15, naslim10, naslim13} and \citet{stroeer07},  and the sample 
subdivided according to the above criteria. 

Six intermediate-helium sdBs have been found to have unusual and interesting surface chemistries. 

\citet{edelmannphd} found PG\,0909+276 and UVO 0512--08 to show surface helium abundances of roughly 20\% by number, 
and extreme (3 -- 4 dex) overabundances of scandium, titanium, vanadium,  manganese, and nickel. 

\lsfour\  has a gravity slightly lower than that of normal sdB stars ($\log g = 5.6$), and a surface helium abundance of about 16 per cent by number. 
It is slightly metal poor ($-$0.8 dex) relative to the Sun. 
What makes  \lsfour\ distinct from any other hot subdwarf, whether helium rich or not, is the  overabundance by 4 orders of magnitude of zirconium, yttrium and 
strontium and a 3 dex overabundance of germanium in the line-forming region of the photosphere \citep{naslim11}.

Two stars, HE 2359--2844 and HE 1256--2738, show optical absorption lines due to triply ionised lead (Pb\,{\sc iv}) which have never previously been detected in 
any star \citep{naslim13}.  The atmospheric abundance of lead is nearly 10 000 times that measured in the Sun. 
HE 2359--2844 also shows zirconium and yttrium abundances similar to those in the zirconium star  \lsfour. 
{UVO 0825+15 has just been discovered to be a variable lead-rich hot subdwarf \citep{Jeffery16}. }

\subsection{\textit{Proper Motions}}
The proper motions used in this work were taken from either the PPMXL or the NOMAD catalogues \citep{roeser10,zacharias03}, 
adopting the value with the smaller error wherever more than one value was available. Proper motions were obtained from \textit{Gaia} DR1  { \citep{gaia2,gaia1,gaia3,gaia4}} for JL\, 87 and SB\, 705.
The mean error of the proper motions is $\pm$3.91 mas/yr.{ Large catalogues such as PPMXL and NOMAD inevitably contain errors and extreme values due to outliers. For this reason the proper motions obtained for this paper were compared with the UCAC4 catalogues; all proper motions agree within errors where the catalogues overlap. { A study of the problems of outliers is discussed in \citet{Ziegerer15}.}

\subsection{\textit{Distances and Reddening}}
Distances were estimated from photometry using the distance modulus 

\begin{equation} \label{eq1}
\mu = 5\log_{10}d - 5 + A_{\rm B },
\end{equation}

\noindent where $\mu$ = $ m_{\rm B }$ - $ M_{\rm B }$ and $A_{\rm B }$ = 4.1$E_{\rm B-V}$. $ M_{\rm B }$ is the absolute magnitude and $ m_{\rm B }$ is the apparent magnitude in B.  Values for $ m_{\rm B }$ were taken from SIMBAD and those for $ M_{\rm B }$ were taken from the literature (references in Appendix Tables \ref{A1}--\ref{A3}). Only 45$\%$  of stars in this sample had published values for $M_{\rm B }$, for the remainder estimates were made by assuming the means of these values, which are 4.14 mag for the helium-rich  and 3.99 mag for the helium-deficient sdO/Bs.
The reddening values ($A_{\rm B }$) were found using a dust extinction tool, hosted by the NASA/ IPAC Infrared Science Archive, which gives the Galactic dust reddening for a line of sight, returning a reddening map that is a reprocessed composite of the COBE/DIRBE and IRAS/ISSA maps \citep{dust}.  {The large reddening corrections to the KPD stars may be overestimated by this method due to their low Galactic latitudes. Hence their distances are likely underestimated.}

Distances, from a variety of methods, were found in the literature. These were compared with the calculations performed here (Table \ref{distance}). Out of the twenty distances compared there was a 95\% agreement.
 {Distances for JL\, 87 and SB\, 705 were taken from the \cite{BJ} catalogue which make use of the $Gaia$ DR1 parallaxes.}
\begin{table}

\caption[]{Calculated distances in kpc compared with published values. } 

\begin{tabular}{|l|c|c|}
\hline
Star  &  Calculated Distance & Published Distance \\
\hline
HE\,0004--2737  & 0.62   $\pm$     0.09&0.67\footnotemark[1]  \\
HE\,0151--3919  &  1.07  $\pm$    0.27 &  0.92\footnotemark[1] \\
HS\,0232+3155   &   1.95 $\pm$ 0.27  &   1.70\footnotemark[1] \\
HS\,0233+3037  &   1.23 $\pm$ 0.17 &   1.00\footnotemark[1]\\
HE\,0407--1956  &  0.86  $\pm$    0.12  & 0.89\footnotemark[1] \\
HS\,0546+8009  &   1.10 $\pm$ 0.15 &   1.00\footnotemark[1]\\
HS\,0815+4243  &  2.60  $\pm$  0.35 &   2.50\footnotemark[1]\\
PG\,1136--003  &  0.87  $\pm$ 0.23 &    1.3 $\pm$ 0.2\footnotemark[2]\\ 
HS\,1236+4754  &   1.99 $\pm$  0.27&  2.10\footnotemark[1]\\
HS\,1320+2622  &  3.14  $\pm$ 0.43 &  3.40\footnotemark[1]\\
HS\,1739+5244  &  1.76  $\pm$  0.24 &   1.80\footnotemark[1]\\
HS\,1741+2133  &   1.95 $\pm$  0.27&  1.80\footnotemark[1]\\
HE\,2135--3749  &  0.66  $\pm$  0.10& 0.71\footnotemark[1]    \\
HS\,2156+2215  &   3.05 $\pm$  0.42&  2.80\footnotemark[1]\\
HS\,2201+2610  &   0.95 $\pm$  0.13& 0.90\footnotemark[1]\\
HS\,2208+2718  &   1.28 $\pm$  0.18&  1.20\footnotemark[1]\\
HS\,2242+3206  &    1.48 $\pm$  0.20&    1.30\footnotemark[1]\\
HE\,2337--2944  &  0.90 $\pm$  0.13&   0.96\footnotemark[1]   \\
CD--35$^{\circ}$15910  &  0.22  $\pm$  0.06 & 0.22 $\pm$ 0.08\footnotemark[3]  \\ 
PG\,2352+181  & 0.38 $\pm$ 0.06 & 0.90 $\pm$ 0.32\footnotemark[4]\\
\hline

\end{tabular}
{  \footnotemark[1] \cite{edelmannphd},  \footnotemark[2] \citet{muchfuss15}, \footnotemark[3] \citet{para07}, \footnotemark[4] \citet{kin94} }

\label{distance}
\end{table}

\section{Kinematics}

\begin{table*}

\caption{{Mean values and standard deviations of the hot subdwarf helium classes. Results from \citet{altmann04} and \citet{pauli06} are also shown here.}}

\begin{tabular}{lccccccc}

\hline
Subsample & N & $\overline{U}$ & $\sigma_U$ &$ \overline{V}$& $ \sigma_V$ &$\overline{W}$&$\sigma_W$\\
\hline
All					&	88	&	--1.7		&	72.6		&	202.2	&	67.6		&	5.6		&	49.0\\
Helium-Deficient		&	38	&	14.1		&	56.7		&	215.0	&	52.9		&	--4.4		&	39.8\\
Intermediate-Helium		&	27	&	3.1		&	83.3	&	190.2	&	84.7	&	13.4		&	35.5\\
Extreme-Helium		&	23	&	--30.4	&	72.6		&	193.8	&	65.9		&	13.9		&	69.9\\

\hline
Altmann &114&--8&74&198&79&12&64\\
Pauli WD thin disk &361&...&34&...&24&...&18\\
Pauli WD thick disk&27&...&79&...&36&...&46\\
LSR &...&0&...&242&...&0&...\\

\hline
\end{tabular}

\label{sd}
\end{table*}

\subsection{Calculating Galactic Velocities}

Using the observed values of right ascension, declination, distance, proper motion and radial velocity, the Galactic velocity components were calculated following the method outlined in \citet{randall15}. The left-handed system for the velocity components is used here, where $U$ is the Galactic radial velocity, positive toward the Galactic centre, $V$ is the Galactic rotational velocity in the direction of the Galactic rotation and $W$ is the component positive toward the North Galactic Pole.
This calculation assumes the distance of the Sun from the Galactic centre to be 8.4 kpc, its motion relative to the Local Standard of Rest (LSR) to have components ($v_x$,$v_y$,$v_z$) = (11.1, 12.24, 7.25)\kmsec\ and the velocity of the LSR to be $V_{LSR}$ = 242\kmsec \citep{irrgang13}.

\begin{figure}
\center
\includegraphics[width = 220px]{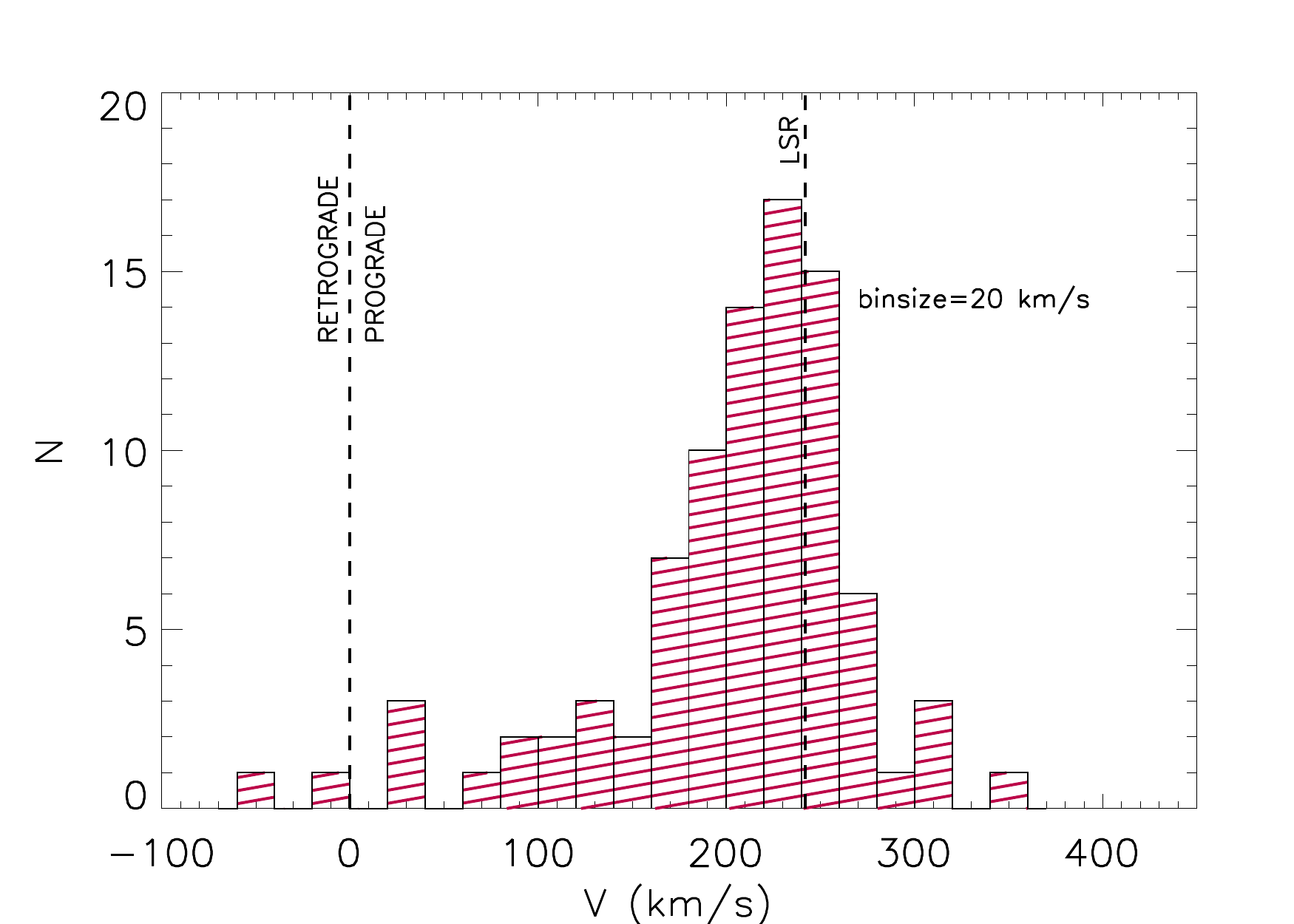}
\caption{{Histogram of the Galactic rotational velocities of all 88 stars of the sample.}}

\label{V_hist}
\end{figure}

\subsection{Galactic velocities and velocity dispersions}

Figure \ref{V_hist} shows the distribution of rotational velocity, $V$, for all 88 stars. 
 {A peak can be seen at 220\kmsec, near the local standard of rest (LSR) where disk stars are expected to be. This has also been found in \citep{pauli06} for their white dwarf sample.}
This distribution is similar to the sample of 114 sdBs analysed by \citet{altmann04} who also found a similar peak and a tail of the distribution extending into negative rotational velocities.

The kinematics of this sample were first investigated using a $U-V$ velocity diagram. 
In Fig. \ref{UV}, the Galactic radial velocity $U$ is plotted versus the Galactic rotational velocity $V$. 
The dotted ellipses correspond to the 3$\sigma$-thick and 3$\sigma$-thin disk distributions of white dwarfs as calculated by \citet{pauli06}, who classified a sample of 398 white dwarfs based on their position in the $U-V$ diagram, the position in the $J_z$ -- eccentricity diagram and the stars Galactic orbit combined with age information.
 {In the $U-V$ diagram (Fig. \ref{UV}) the helium-deficient stars (green) are mostly clustered around the LSR (black star).  }
The helium-rich stars (intermediate and extreme) are more widely distributed in this plot. 
Two of these stars have possible retrograde orbits, with their Galactic rotational velocities being negative, they are labelled in Fig.~\ref{UV}.
HE\,2218--2026 lies far outside the disk distributions with a high Galactic radial velocity of almost 300\kmsec.

Another method of analysing the kinematics of stars is to look at their total velocity or kinetic energy. In Fig. \ref{kinetic} the kinetic energy $2E_{\rm kin}/m =U^2 + V^2 +W^2$ is plotted against the rotational velocity $V$. 
Included in this graph are contours showing the velocities perpendicular to Galactic rotation at certain values, where $v_{\bot} = (U^2 + W^2)^{1/2}$.
The higher the value of $2E/m$ the more an object deviates from a circular orbit. For low values of $v_{\bot}$ the deviation from LSR gives information about the kinetic temperature. 
The clustering of values around $v_{\bot}$ = 0 means that they are kinematically cool. 
A few stars are located further away from the $v_{\bot}$ = 0 contour;  these are the kinematically hot stars and likely to have a more eccentric orbit. Another reason for a large proportion of stars with low $v_{\bot}$ could be that they are near their orbital turning point.
Table \ref{sd} shows the mean values and standard deviations of the galactic velocities for the hot subdwarf helium classes as compared with previous studies.
The Galactic velocities and corresponding errors for individual stars are shown in appendix Tables \ref{heinter} -- \ref{hedef}.

\section{Galactic Orbits}

\subsection{Calculating the orbits}
In addition to Galactic velocities the orbits were calculated for the stars in this sample. This was achieved using \verb|galpy|, a python package for Galactic-dynamic calculations \citep{bovy15}. The orbits were computed using the potential MWPotential2014, this model is fit to dynamical data of the Milky Way. Although this is not the best possible current model, it was chosen as it gives a realistic model of the Milky Way's gravitational potential that is simple and easy to use. It consists of a bulge modelled as a power-law density profile that is exponentially cut off with a power-law exponent of --1.8 and a cut-off radius of 1.9\,kpc, a Miyamoto-Nagai Potential disk \citep{Miy75}, and a dark-matter halo described by a Navarro-Frenk-White potential \citep{Navarro96}.  {The distance of the Sun from the Galactic centre is set to 8.4 kpc and the velocity of the LSR is $V_{\rm LSR}$ = 242\kmsec.}

Extracted from these orbits, integrated over $\approx$ 3Gyrs, are the quantities apocentre, $R_{\rm a}$, pericentre, $R_{\rm p}$, eccentricity, $e$, maximum vertical amplitude, $z_{\rm max}$, and normalised z-extent, $z_{\rm n}$. The quantities $R_{\rm a}$ and $R_{\rm p}$ are the maximum and minimum distances from the Galactic centre attained during a revolution of 2$\pi$ radians, measured on the Galactic plane. From these distances we find the eccentricity which is given by
\begin{equation}
e= \frac{R_{\rm a} -R_{\rm p}}{R_{\rm a}+R_{\rm p}}
\end{equation}
The normalised z-extent of the orbit, which can be used as a measure for the inclination of an orbit, is given by
\begin{equation}
z_n = \frac{z_{\rm max}}{R(z_{\rm max})}
\end{equation}
where $R$ is the galactocentric distance. These quantities are shown in appendix Tables \ref{heinter}--\ref{hedef}, along with the star's Galactic velocities. The mean values and standard deviations for the parameters $e$, $z_{\rm n}$ and $z_{\rm max}$ are shown in Table \ref{orbit}. 

{Two important orbital parameters are the z-component of the angular momentum $J_z$ and the eccentricity of the orbit. Fig. \ref{jze} shows a plot of $J_z$ versus eccentricity. This diagram can be used to distinguish different populations. The thin disk stars cluster in an area of low eccentricity and $J_z$ around 1800 kpc\kmsec. \cite{pauli03} call this Region 1. The thick disk stars possess higher eccentricities and lower
angular momenta called Region 2. Those stars which lie outside these regions are halo candidate stars.}

\begin{figure}
\center
\includegraphics[width = 220px]{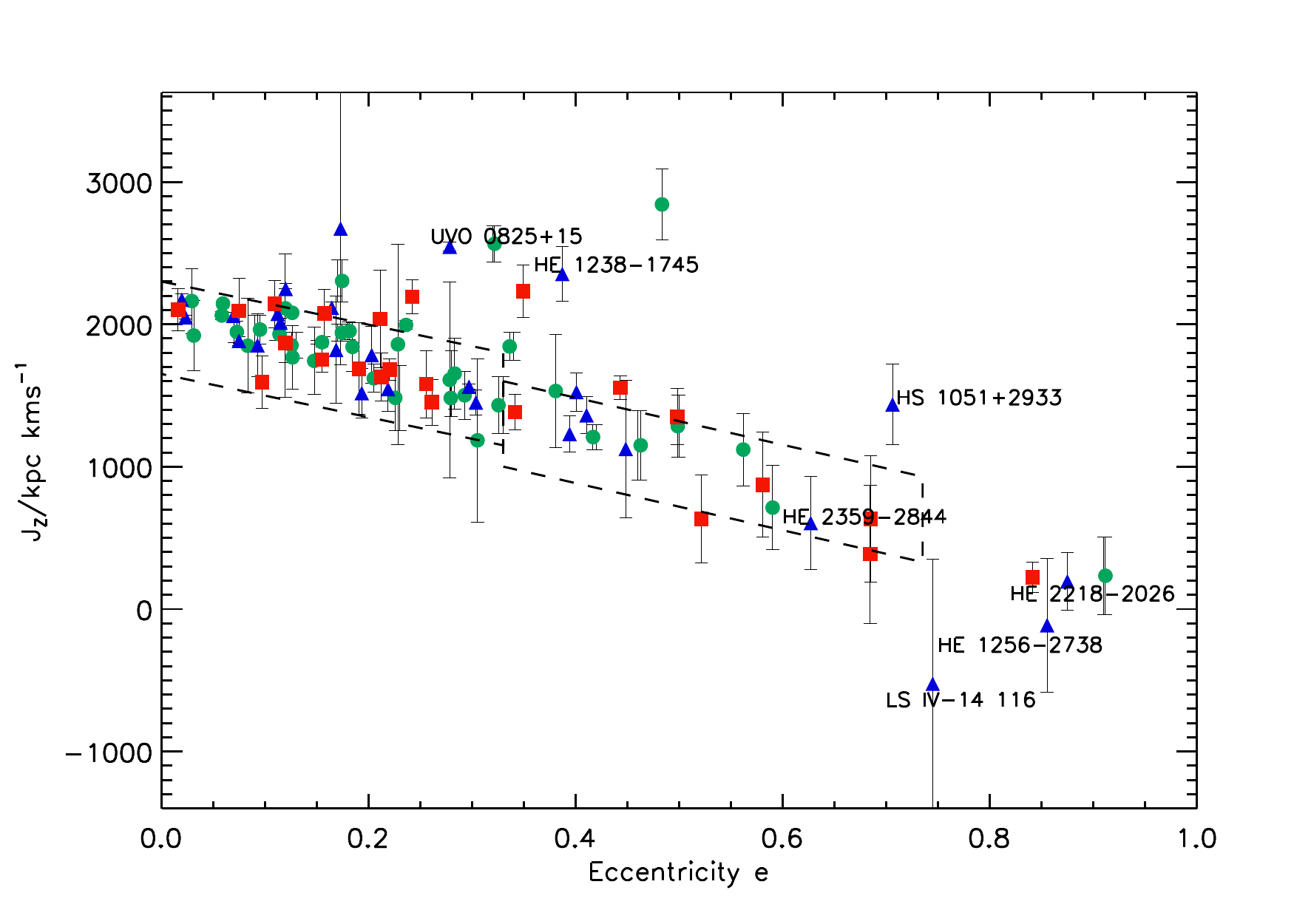}
\caption{{Z-component of the angular momentum versus eccentricity. Symbols have same meaning as previous plots. Eccentricity errors have been removed for clarity.}}
\label{jze}
\end{figure}

\begin{figure*}
\center
\includegraphics[width = 17cm]{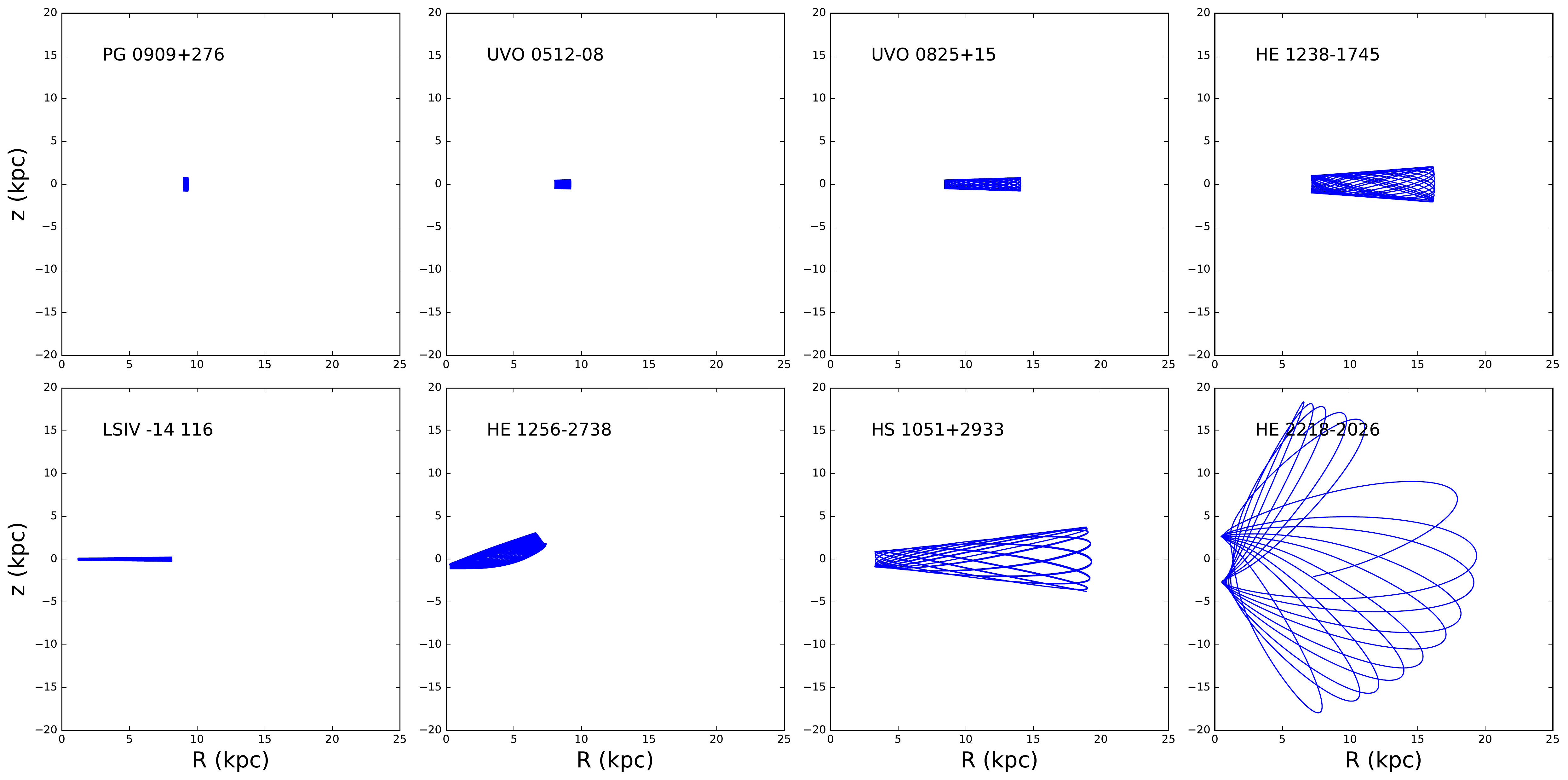}
\caption{Meridional sections of the orbits of 8 stars. Columns 1$-$2 in row 1 are examples of probable thin disk stars and columns 2$-$4 include probable thick disk orbits. Row 2 contains probable halo orbits.}
\label{meridional}
\end{figure*}

\begin{figure*}
\center
\includegraphics[width = 17cm]{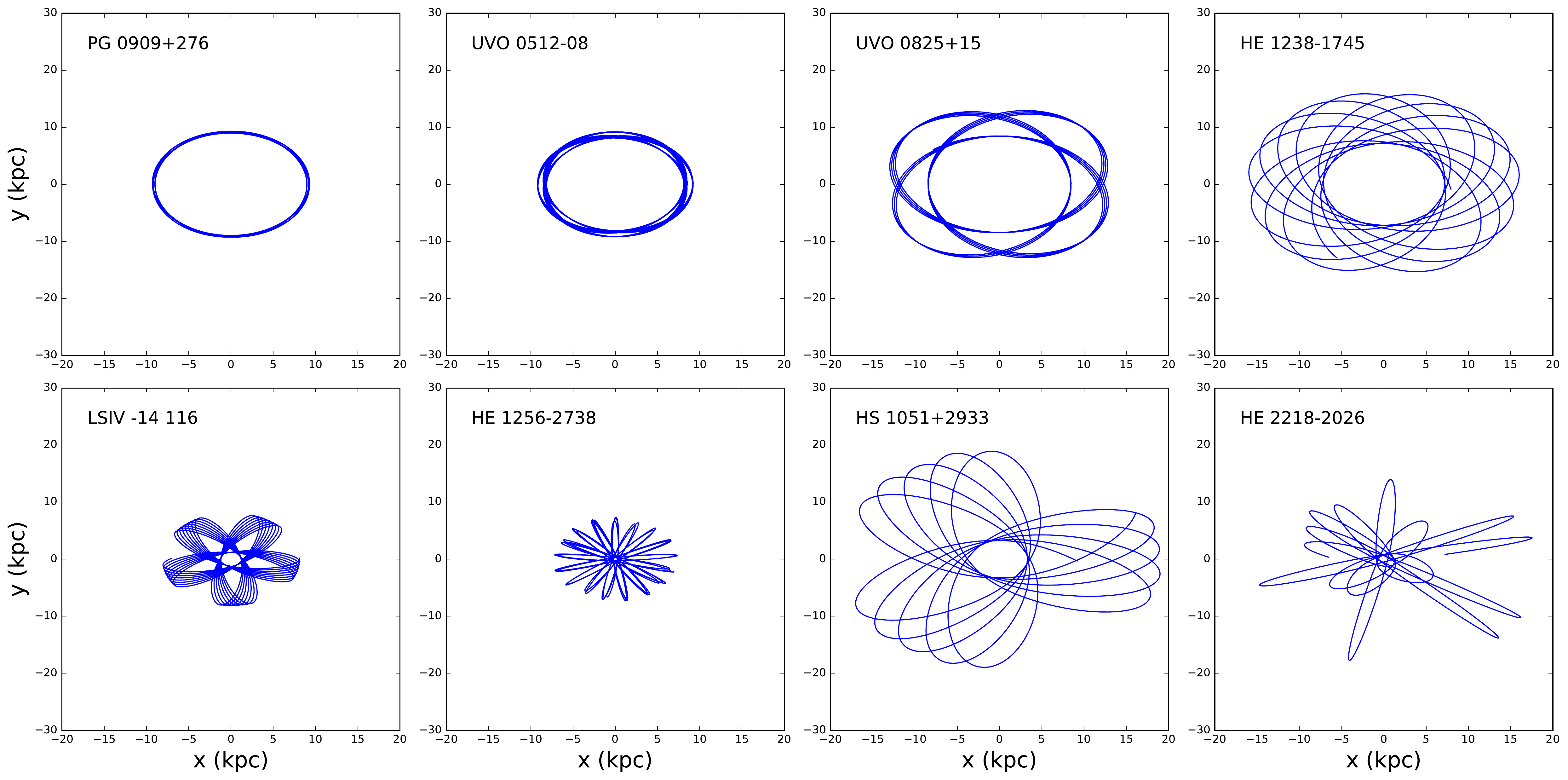}
\caption{As Fig. \ref{meridional}, but projected on the x-y Galactic plane.  }
\label{xy}
\end{figure*}

\subsection{The orbits and orbit parameters}

\begin{table*}

\caption{{Mean values and standard deviations of the orbital parameters eccentricity, maximum z--extent, maximum vertical amplitude, apocentre and pericentre, respectively.}}
\begin{tabular}{cccccccccccc}

\hline
Subsample & N &$\overline{e}$&$\sigma_{e}$&$\overline{z_{\rm n}}$&$\sigma_{z_{\rm n}}$&$\overline{z_{\rm max}}$&$\sigma_{z_{\rm max}}$& $\overline{\rm R_{a}}$ &$\sigma_{\rm R_{a}}$& $\overline{\rm R_{p}}$&$\sigma_{\rm R_{p}}$\\
\hline
All	&88&	0.29	&0.22&	0.23&0.35	&1.76&	2.55&	10.03&	2.85&	5.70	&2.26\\
Helium-Deficient&	38&	0.26	&0.18&	0.16	&0.17&	1.19&	1.18&	9.91&	2.75	&5.99&	1.97\\
Intermediate-Helium	&27&	0.32&	0.23	&0.27&	0.41	&1.92&	3.02	&10.31&	2.78	&5.49&	2.39\\
Extreme-Helium	&23&	0.32	&0.23&	0.31&	0.34&	2.53	&2.78&	9.91&	2.57	&5.44&	2.44\\

\hline
\end{tabular}

\label{orbit}
\end{table*}
 
\begin{figure}
\center
\includegraphics[width = 250px]{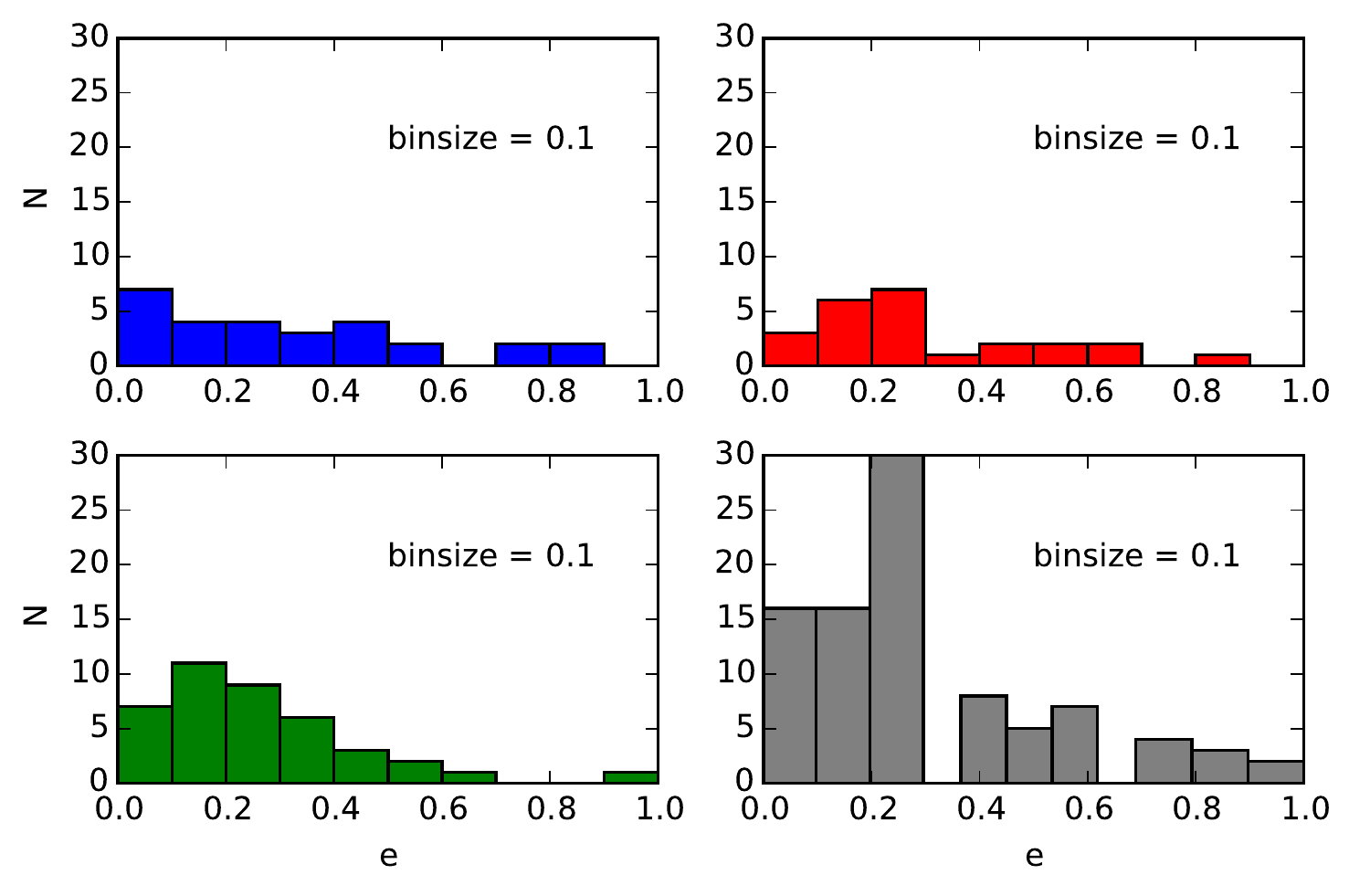}
\caption{{Histograms of the eccentricity distribution. Top left: intermediate-helium, top right: extreme-helium, bottom left: helium-deficient, bottom right: all.}}
\label{nze_hist}
\end{figure}

Fig. \ref{nze_hist} shows histograms of the eccentricity distribution of the three helium classes. The He-deficient and total group of stars has a peak of $e$ $\approx$  0.2, this region is mainly populated by disk stars. The He-enriched stars, on the other hand, have a distribution widely spread over the whole range.

{Our classification scheme has been adopted from \cite{pauli03,pauli06}.  {\citet{Ma16} found a vertical scale height for the thick disk between 1 and 1.5\ kpc, so here we have set 1.5\ kpc as the cut-off height for the thin disk.}}}

\begin{enumerate}

\item Stars whose orbits have low eccentricity and low inclination on the Galactic plane are referred to as the thin disk component. 
These stars must lie within region 1 in the $J_z$ -- $e$ diagram and within the 3$\sigma$-thin disk contour and have $z_{\rm max}\ <$ 1.5\ kpc

\item {Stars which lie in Region 2 and the 3$\sigma$-thick disk contour have been classified as belonging to the thick disk.}
\item {Stars which have been classified as halo lie outside Regions 1, 2 and lie outside both 3$\sigma$ contours and have a $z_{\rm max}\ >$ 1.5\ kpc.}\\
{ {All orbits have also been visually inspected as part of the classification.}}

\end{enumerate}

Table \ref{pop} shows the number of stars classified as halo, thin or thick disk stars.
The thin and thick disk kinematically overlap one another in a way which makes it difficult to find selection criteria capable of distinguishing them from one another. 
It has been suggested by \citet{bovy11} that there is actually no thin/thick disk dichotomy and that the transition between thin and thick disk is rather a continuum of disks.

There is a large variety of orbit morphologies in this sample.
Fig. \ref{meridional} shows the meridonial sections of the orbits of 18 stars. 
Row 1 contains intermediate helium stars, row 2 contains extreme helium sdO/Bs and row 3 shows the orbits of the helium-deficient sdO/Bs. 
Columns 1--3 are examples of disk stars and columns 4--6 show halo orbits.
 The same layout applies to Fig. \ref{xy} which shows the orbits projected onto the $x-y$ Galactic plane. 

\begin{table}
\caption{{Population classification}}
\begin{tabular}{lcccc}

\hline
Subsample & N &Thin Disk&Thick Disk&Halo\\
\hline
All					&	88	&	45	&	36	&	7\\
Helium-Deficient		&	38	&	25	&	12	&	1*\\
Intermediate-Helium		&	27	&	13	&	10	&	4\\
Extreme-Helium 		&	23	&	7	&	14	&	2\\

\hline

\end{tabular}
{* The halo classification for the helium-deficient HS\, 2208+2718 should be considered provisional,  as there is a non-zero probability that it is a radial velocity variable.}
\label{pop}
\end{table}

\section{Discussion}

In general, hot subdwarfs show a much larger distribution in velocity than white dwarfs. 
Figure \ref{UV} shows that the helium-deficient subdwarfs are more clustered around the LSR which
suggests that they are likely disk stars with some having possible thin disk kinematics. 
Two of the chemically--peculiar sdBs have a very low or even retrograde
rotational velocity indicating that they have halo kinematics.
The kinetic energy diagram (Fig. \ref{kinetic}) shows a significant portion of the sample have kinematics that are solar-like
suggesting membership of the thin disk. 
The mean values and standard deviations of the galactic velocities of the entire sample of sdO/Bs agree amiably with Altmann's 
study of 114 sdBs.
Examining the helium-deficient  velocity distribution alone (Table \ref{sd}), it may be seen that the classical
sdB stars  fit closer to Pauli's white-dwarf thick-disk sample than to Altmann's sample. 
The helium-rich groups have much higher standard deviations (Table \ref{sd}), suggesting that there is more of a 
halo contribution to these subgroups.
{ \cite{kawka15} calculated the Galactic velocity components of all known hot subdwarf binary systems. They found that the population kinematics imply an old age and include a few likely halo population members and that the hot subdwarf binary population has a velocity dispersion between the thin and thick disk dispersions for white dwarfs. }

Special attention has been paid to intermediate-helium stars with peculiar surface chemistries  (Figs. \ref{meridional} and \ref{xy}). \\
\noindent{\bf LS\,IV--14$^\circ$116} (Zr, Y, Sr, Ge) initially appears to have a disk orbit due to its maximum vertical 
amplitude ($z_{\rm max}$ = 0.26kpc), very low inclination and the shape of its meridional 
orbit (Fig. \ref{meridional}). 
But due to its high eccentricity ($e$=0.74) and retrograde velocity ($V=-54$\kmsec) it clearly has a halo orbit; cf. \citet{randall15}. \\
\noindent{\bf HE\,1256--2738} (Pb) immediately presented as a halo orbit. 
It has a very low galactic rotational velocity and a very high eccentricity of 0.86. 
It also has a very close approach to the galactic centre with a pericentre value of 0.59\,kpc,  
which could account for its deviation from a circular orbit. \\
\noindent{\bf HE\,2218--2026},  an intermediate-helium hot subdwarf, stands out as having a halo orbit 
with a large radial, low rotational velocity and a chaotic orbit
(Fig. \ref{meridional}) which closely approaches the galactic centre. All of its orbital parameters
are characteristic of the halo. At 18\,kpc, it has the highest $\rm z_{max}$ of any of the 
stars in this sample. The Galactic rest-frame velocity of 321\kmsec\ is not high enough to be classified as a hyper-velocity star, 
{\it i.e.} as having rest-frame velocity greater than that of the local Galactic escape velocity ($\approx$ 500\kmsec), cf. the review by \cite{brown15}.

\noindent{{UVO\,0825--08 (Ca, Ge, Y, Pb), HE\, 1238--1745 and HS\, 1051+2933 are all marginally outside the thick disk 3$\sigma$ contour and have $\rm z_{max}$ =  0.86, 2.08 and 3.76 kpc respectively. HS\, 1051+2933 is clearly a halo star, the other two are marginally halo/thick disk stars. We have been conservative in assigning stars to the lower energy population.              }}
 
\noindent{\bf PG\,0909+276 and UVO\,0512--08} (Sc, Ti, V, Mn, Ni), although chemically interesting intermediate-helium  subdwarfs, appear to show typical thin disk orbits. \\ 

These results are helpful for interpreting the evolutionary origin of all types of hot subdwarfs. For the 
helium-deficient or \textquotedblleft normal\textquotedblright\ sdB stars, an origin in a close binary system is considered likely for the majority;
indeed, 15/38 of the sample are confirmed binaries. Whether these systems originate in one or more 
common-envelope ejection episodes, or in a stable Roche-lobe overflow episode, the age and mass of the progenitor are closely 
linked by the progenitor main-sequence lifetime. Models imply binary sdB progenitors may have masses in the range $0.8 < M_{\rm MS}/\Msolar < 5$
or more \citep{han1}, implying possible ages anywhere between 0.2 and 10 Gyr, and hence an origin in either disk or halo. 
Significantly, the fractions for both classes of helium-rich subdwarfs are much lower, only one intermediate-helium subdwarf, CD--20$^\circ$1123, is a confirmed close binary \citep{naslim12}. 

For extreme-helium subdwarfs, \citet{zhang12} argue for an origin in a merging double-helium white dwarf 
binary. 
Such systems require a significant delay between formation of the double white dwarf and the subsequent merger because, for these stars, 
orbital decay by gravitational radiation has a timescale $\geq 1$\,Gyr.  
Binary star population synthesis studies show that double helium-white dwarf progenitor systems  must have formed at least 2\,Gyr ago, 
with 95\% formed more than 4\,Gyr ago \citep{yu11,zhang14}. One would therefore expect kinematics representative of an older population,  
as suggested by Table~\ref{pop}. 

The intermediate-helium subdwarfs present more of a challenge.  
\citet{naslim11,naslim13} have argued that the extreme surface abundances seen in \lsfour\ and the lead stars HE\,2359--2844 and 
HE\,1256--2738 are proto-subdwarfs, evolving {\it onto} the extended horizontal branch. 
Their atmospheres should consequently represent a snapshot of an evolving surface chemistry in which radiative levitation 
and gravitational settling continually resort the surface layers as the star evolves  and helium sinks out of the photosphere. 
For this to be true, the intermediate-helium subdwarfs should share the kinematical properties 
of the helium-deficient subdwarfs. 
 {If the fraction of halo stars in each subsample (Table \ref{pop})  is representative of age, then the intermediate-helium subdwarfs 
{(4 halo / 27 stars) would appear to be older than both the helium-deficient (1/38) and the extreme-helium subdwarfs (2/23)}. However,  the statistics are at best small. Radial velocities have not been obtained in a uniform manner so that a high fraction of extreme-helium subdwarfs have only single epoch data, whilst the other groups are represented by multi-epoch data. Similarly, the sample space volumes are slightly different, with mean distances for the three groups being 
$\langle d\rangle=1.13\pm0.89$\,kpc (intermediate), 
$\langle d\rangle=1.39\pm0.90$\,kpc (extreme), 
and
$\langle d\rangle=1.07\pm0.79$\,kpc (helium-deficient). 
This could account for the slightly lower energy classification of the helium-deficient group with respect to 
the helium-rich groups. 
Moreover, our sample of helium-deficient subdwarfs omits a more distant sample of halo sdBs. \citet{tillich11} identified two distinct kinematic groups: normal halo subdwarfs with low Galactic rotation and extreme halo subdwarfs on highly-eccentric retrograde orbits. The presence of two distinct groups indicates different origins. The normal halo sdBs might have been ejected into the halo via the slingshot mechanism. The extreme halo stars might originate in the outskirts of the Galactic disc and not in the central bulge. It is therefore not unreasonable to suspect that the intermediate-helium subdwarfs equally show such a diverse range of kinematical properties.}

However, it is puzzling that the two intermediate-helium subdwarfs with the most extreme chemistries are 
in  halo orbits. An intriguing alternative is that these stars represent the ejecta from Type Ia supernovae, as suggested for 
the hyper-velocity compact helium star US\,708 \citep{justham09,geier13}. A hot subdwarf having a massive white dwarf 
companion in a short-period orbit, such as CD$-30^{\circ}11223$, will, towards the end of core helium burning, expand 
and  transfer mass to its companion, potentially stripping the hydrogen layers from the subdwarf and leading to a 
thermonuclear explosion in the white dwarf. The explosion will have the consequences of i) contaminating 
the helium-rich subdwarf remnant with heavy-metal ejecta from the supernova, and ii) unbinding it with a velocity
close to the orbital velocity at the time of explosion $\approx 300-500$\kmsec\ \citep{liu13}. The result is a chemically-peculiar
helium-rich subdwarf in a halo-like orbit. Whether bound or unbound will depend on the subdwarf velocity vector relative
to the Galactic potential at the time of explosion.

\section{Conclusion}
  
The space motions and Galactic orbits of 88 hot subdwarfs were computed from published proper motions,  radial velocities and inferred distances. 
The orbital parameters were used to classify sample members as having disk or halo orbits. 
This study confirms that sdO/B stars are members of all Galactic populations. 
The sample was divided into helium-deficient (or {\it normal}), intermediate-helium and extreme-helium subdwarfs based on their surface helium abundances,
in order to establish whether the different  groups could be distinguished kinematically.
 {Of the samples studied, helium-deficient sdO/Bs show the lowest standard deviations in all orbital parameters discussed here
and are likely to be primarily disk stars. However at least one and possibly two populations of halo sdB stars are also known to exist \citep{tillich11}. 
Both samples of He-rich subdwarfs appear to have similar  kinematics, primarily comprising disk stars with a small fraction of halo objects.}

Three intermediate-helium stars, including two with peculiar surface chemistries, show quite extreme halo orbits, but their space velocities
are insufficient for them to be {\it bona fide} hyper-velocity stars. The high velocities argue {\it against} a previously proposed connection 
between the chemically-peculiar and normal hot subdwarfs, but pose equally challenging questions concerning the origin of the former. 
The possibility that they are the polluted ejecta from Type Ia supernovae appears to be worth exploring further.    

The primary limitation of this study is the small sample size, particularly with regard to the extreme- and intermediate-helium subsamples.
In the near future, the {\it Gaia} spacecraft will deliver distances and proper motions having two orders of magnitude better precision than 
currently available; many of the questions raised here will be addressed within a very few years.

\section*{Acknowledgments}
This research has made use of the SIMBAD database, operated at CDS, Strasbourg, France and is also based on data 
obtained from the ESO Science Archive Facility.\\
We are grateful to James Murphy, then of St Malachy's College, Belfast,  who during 2012 July 
first made the measurements shown in Table 1 whilst holding a Nuffield Science Bursary at the Armagh Observatory .  
The Armagh Observatory is funded by direct grant from the Northern Ireland Dept. for Communities.

 {This work has made use of data from the European Space Agency (ESA)
mission {\it Gaia}\footnote{${http://www.cosmos.esa.int/gaia}$}, processed by
the {\it Gaia} Data Processing and Analysis Consortium (DPAC)\footnote{${http://www.cosmos.esa.int/web/gaia/dpac/consortium}$}. Funding
for the DPAC has been provided by national institutions, in particular
the institutions participating in the {\it Gaia} Multilateral Agreement.}

\bibliographystyle{mnras}
\bibliography{mybib}

\section*{Appendix}
\renewcommand\thetable{A.\arabic{table}} 

The input data, including radial velocities, proper motions and inferred distances for each of the three subdwarf groups 
are presented in Tables \ref{A1} -- \ref{A3}. The orbital parameters, space motions, and inferred population are presented in 
Tables \ref{heinter} -- \ref{hedef}. The table contents are described fully in the text (\S\S 2 and 3). 

%\tiny
\scriptsize

\begin{table*}
\begin{minipage}{180mm}
\caption[]{Input data for the intermediate helium-rich stars. { \footnotemark[2] highlights confirmed binary systems in which case the system velocity $\gamma$ is quoted. * corresponds to single epoch radial velocity measurements. \footnotemark[3] marks distances obtained from {\cite{BJ}.}}}
\centering
\vspace*{0.3cm}

\begin{tabular}{lccccccccccc}
\hline

Star & RV & $\pm$ & ref & $\mu_{\alpha}$ & $\pm$ & $\mu_{\delta}$ & $\pm$ &ref& $M_{\rm B }$ & $d$ &$\pm$\\
%Star&RV &$\pm$&ref&pmRA &$\pm$&pmDEC&$\pm$&ref&	$ M_{\rm B }$	&d &$\pm$\\
       &\kmsec   &&&mas\,yr$^{-1}$&&mas\,yr$^{-1}$&&&&kpc&\\
\hline
UVO 0512--08						&	11.0		&	3.3	& 	$a$	&	--28.5	&	1.3	&	--24.0	&	2.9	&	PPMXL	&	...	&	0.22		&	0.03\\ %?
BPS CS 22946--0005					&	--57		& 10	&	$b$	&	3.2		&	4.1	&	--3.5	&	4.1	&	PPMXL	&	...	&	1.30		&	0.21\\ %y
BPS CS 22956--0094					&	--21		&10		& 	$b,t$	&	37.3		&	2.8	&	--27.5	&	2.8	&	NOMAD	&	...	&	0.59		&	0.09\\ %y
CPD--20$^{\circ}$1123\footnotemark[2]	&	--6.3	&	1.2	& 	$c$	&	6.6		&	1.7	&	--14.8	&	3.6	&	NOMAD	&	...	&	0.29		&	0.05\\ %y
HD 127493							&	7		&	3	& 	$d$$e$&--32.9	&	1.3	&	--16.6	&	1.2	&	PPMXL	&	...	&	0.11		&	0.02\\ %y
HE 1135--1134						&	24.67	& 2.45	&	$t$&	--9.4	&	6.0	&	2.3		&	6.0	&	PPMXL	&	3.88$f$	&2.15	&	0.30\\ %y
HE 1136-2504						&	59.39	& 9.22	&	$t$&	--5.1	&	3.0	&	--6.6	&	2.8	&	NOMAD	&	4.25$f$	&0.89	&	0.12\\%y
HE 1238-1745						&	-7.87	&  2.72	&	$t$&	12.4		&	5.0	&	3.1		&	5.0	&	PPMXL	&	3.83$f$	&1.33	&	0.18\\%y
HE 1256-2738						&	140.46	& 5.6	&	$t$&	--11.3	&	6.6	&	--10.6	&	6.6	&	PPMXL	&	4.04$f$	&3.16	&	0.49\\%y
HE 1310--2733						&	41.54	&  1.82	&	$t$&	--6.5	&	4.6	&	0.6		&	4.6	&	NOMAD	&	3.76$f$	&1.39	&	0.19\\%y
HE 2218--2026						&	--278.86	&  1.45	&	$t$&	20		&	4	&	--4		&	1	&	NOMAD	&	4.40$f$	&2.50	&	0.35\\%y
HE 2357--3940						&	--18.38	&14.15	& 	$g$&	15.8		&	1.2	&	3.7		&	1.2	&	NOMAD	&	...	&	0.15		&	0.02\\%y
HE 2359-2844						&	--90.71	& 3.03	&	$t$&	2		&	3	&	--12		&	3	&	NOMAD	&	3.85$f$	&2.42	&	0.33\\ %y
HS 1051+2933						&	--130	& 3		&	$h$	&	14		&	3	&	--16		&	3	&	NOMAD&	4.70$h$	&1.96	&	0.27\\%y
JL 87								&	--6.1	& 2.3	&	$i$	&	--0.4	&	1.06	&	3.77		&	1.49	&	GAIA&	...	&	0.58		&	0.14\footnotemark[3]\\%y
\lsfour								&	--150	& 2		&	$j$	&	9.2		&	1.8	&	--130.6	&	1.8	&	PPMXL	&	...	&	0.44		&	0.20\\%y
PG 0229+064						&	7.6		& 4.0	&	$k$	&	--16.0	&	1.9	&	--2.0	&	1.9	&	PPMXL	&	...	&	0.33		&	0.05\\%y
PG 0240+046						&	63.4		& 2.0	&	$k$	&	28.4		&	2.7	&	--5.4	&	2.6	&	PPMXL	&	...	&	0.77		&	0.12\\%y 
PG 0909+276						&	20.0		& 2		&	$h$	&	0.5		&	1.0	&	0.1		&	0.9	&	NOMAD	&	...	&	0.87		&	0.14\\ %y
SB 705							&	4		& 12	&	$a$	&	10.36		&	0.53	&	9.84		&	0.61	&	GAIA	&	...	&	0.67		&	0.16\footnotemark[3]\\ %y
TON 107							&	28.5		& 2.8	&	$l$	&	--2		&	6	&	--14		&	5	&	NOMAD	&	...	&	2.08		&	0.34\\%y

UVO 0825+15						&	56.4*	& 0.5		&	$t$	&	--23.7	&	1.2	&	--0.2	&	1.2	&	NOMAD	&	...	&	0.33		&	0.05\\%
SDSS J092440.11+305013.16			&	2.7*	& 0.5		&	$t$	&	0.2		&	4.5	&	--10.4	&	4.5	&	PPMXL	&	...	&	1.01		&	0.16\\%
SDSS J160131.30+044027.00			&	-26.7*	& 0.9		&	$t$	&	--14.8	&	4.3	&	0.2		&	4.3	&	PPMXL	&	...	&	0.89		&	0.14\\%
SDSS J175137.44+371952.37			&	-73.6*	& 0.2		&	$t$	&	--9.3	&	4.2	&	0.5		&	4.2	&	PPMXL	&	...	&	1.32		&	0.21\\%
SDSS J175548.50+501210.77			&	-62.7*	& 0.2		&	$t$	&	--9.9	&	1.6	&	25.2		&	1.6	&	PPMXL	&	...	&	0.42		&	0.07\\%
HS 1000+471 						& 0*	&10		&$t$		&--1.5		&5.6	&--5.8&5.6&PPMXL&...&4.14&0.65\\

\hline

\end{tabular}\\
{$a.$ \cite{kilkenny89}, $b.$ \cite{beers92}, $c.$ \cite{naslim12}, $d.$ \cite{bobylev07}, $e.$ \cite{kharchenk07}, 
$f.$ \cite{stroeer07}, $g.$ \cite{kordopatis13}, $h.$ \cite{edelmannphd}, $i.$ \cite{ahm07}, $j.$ \cite{randall15}, 
$k.$ \cite{aznar02}, $l.$ \cite{lou15}, $t.$ This paper: Table 1  }
\label{A1}
\end{minipage}
\end{table*}

%--------------------------------

\begin{table*}
\begin{minipage}{180mm}
\caption[]{Input data for the extreme helium--rich stars. Symbols as in A.6.}
\centering
\vspace*{0.3cm}

\begin{tabular}{lccccccccccc}
\hline

Star & RV & $\pm$ & ref & $\mu_{\alpha}$ & $\pm$ & $\mu_{\delta}$ & $\pm$ &ref& $M_{\rm B }$ & $d$ &$\pm$\\
       &\kmsec   &&&mas\,yr$^{-1}$&&mas\,yr$^{-1}$&&&&kpc&\\
%Star&RV &$\pm$&pmRA &$\pm$&pmDEC&$\pm$&ref&	$ M_{\rm B }$	&d &$\pm$\\
%&        &\kmsec  &&(mas/yr)&&(mas/yr)&&&&(kpc)&\\
\hline
BPS CS 22940--0009			&	37	& 	10		&	$a,t$	&	2.8		&5.2	&	--14.4	&5.2	&	NOMAD	&	...		&	0.77		&0.12\\%y
BPS CS 29496--0010			&	--39.8	& 	0.1		&	$t$	&	--19.0	&4.3	&	--1.2	&4.3	&	PPMXL	&	...		&	1.11		&0.17\\%y
HE 0001--2443				&	3.98		&	0.71		&	$t$	&	6		&2		&	--26		&3		&	NOMAD	&	4.62$i$	&	0.66		&0.1\\%y
HE 0342-1702 				&	--15*		& 	10		&	$a$	&	--3.0	&5.6	&	--8.8	&5.6	&	PPMXL	&	...		&	0.73		&0.10\\%y
HE 1251+0159				&	3* 		& 	24		&	$b$	&	8		&2		&	--18		&3		&	NOMAD	&	4.47$i$	&	1.55		&0.22\\%
LB 3229						&	24		&	 28		&	$c,t$	&	16.2		&5.2	&	--4.4	&5.3	&	NOMAD	&	...		&	0.66		&0.10\\%y
PG 0039+135				&	--92*		& 	66		&	$d$	&	--2.6	&4.9	&	1.8		&5.4	&	NOMAD	&	...		&	0.33		&0.05\\%y
PG 1413+114				&	23.2		& 	17		&	$e$ &	--1.4	&4.6	&	--16.8	&4.6	&	PPMXL	&	...		&	1.99		&0.31\\%
PG 1536+690				&	--295	& 	15		&	$f$	&	0		&2		&	--20		&4		&	NOMAD&	...		&	0.96		&0.15\\
PG 2321+214				&	--19.7*	& 	3		&	$t$ &	23.9		&4.6	&	--8.9	&4.9	&	NOMAD&	...		&	0.49		&0.08\\
PG 2352+181	&	--49*		& 	38		&	$d$ &	26.3		&4.9	&	--3.4	&5.2	&	NOMAD	&	...		&	0.38		&0.06\\
PG 0902+057	&--15*	&5	&$t$&	--15.4	&4.4	&--6.3	&4.4	&PPMXL&	...&	0.93	&0.15\\
PG 1615+413	&--80*	&5	&$t$	&--6.9	&5.6	&--13.5&	5.6	&PPMXL	&...	&2.88&	0.45\\
PG 1600+171	&--78*	&5	&$t$	&--10.5&	5.2	&--4.9&	5.2	&PPMXL&	...	&3.14&	0.47\\
PG 1658+273&	--33*	&5	&$t$&	--19.2	&4.4	&--0.7	&4.4	&PPMXL	&...	&2.33	&0.37\\
PG 1715+273&	31*	&5	&$t$	&--2.0	&3	&--6.0&	1	&NOMAD	&...	&2.91	&0.46\\
HS 1844+637&	20*	&10	&$t$	&--5.0	&5.6	&1.3	&5.6&	PPMXL	&...	&2.7	&0.42\\
PG 1554+408&	73*	&8	&$t$	&--5.1	&5.4&	--3.3	&5.4	&PPMXL&	...	&2.34	&0.37\\
PG 2258+155&	34*	&5	&$t$	&0.6	&5.3&	--5.6	&5.3	&PPMXL&	...	&1.65	&0.26\\
PG 1127+019&	19*	&2	&$t$	&--7.1&	4.1&	--9.2	&4.1&	PPMXL	&...	&0.62	&0.01\\
PG 1415+492&	54*	&1	&$t$	&--1.3&	4&	--3.8	&4	&PPMXL&	...	&1.17&	0.18\\
PG 2215+151&	--13*	&5	&$t$	&3.1&	5.5&	15.1&	5.1	&NOMAD&	...&	0.96	&0.15\\
PG 1544+488\footnotemark[2]&	--25.5	&0.4&	$t$&	--44&	3	&34&	1&	NOMAD&	...	&0.55&	0.09\\

\hline
\end{tabular}\\
{$a.$ \cite{beers92}, $t.$ This paper: Table 1, $b.$ \cite{adelman08}, $c.$ \cite{kilkenny89}, $d.$ \cite{brown08}, 
 $e.$ \cite{muchfuss15}, $f.$ \cite{heber96}, $i.$ \cite{stroeer07}  }
\label{A2}
\end{minipage}
\end{table*}

%---------------------------------
%\renewcommand{\thefootnote}{\arabic{footnote}}

\begin{table*}
\begin{minipage}{180mm}
\caption[]{Input data for the helium--deficient stars. Symbols as in A.6 and A.7.}
\centering
\vspace*{0.3cm}
\label{}
\begin{tabular}{lccccccccccc}
\hline

Star & RV & $\pm$ & ref & $\mu_{\alpha}$ & $\pm$ & $\mu_{\delta}$ & $\pm$ &ref& $M_{\rm B }$ & $d$ &$\pm$\\
       &\kmsec   &&&mas\,yr$^{-1}$&&mas\,yr$^{-1}$&&&&kpc&\\
%Star&RV &$\pm$&pmRA &$\pm$&pmDEC&$\pm$&ref&	$ M_{\rm B }$	&d &$\pm$
%&        &\kmsec   &&(mas/yr)&&(mas/yr)&&&&(kpc)&\\
\hline
CD --35$^{\circ}$15910			&	25.7*		& 	2.8	&	$a$	&	--21.4	&	1.8	&	--6.2	&	1.5	&	PPMXL	&	...		&	0.22		&	0.06\\%yes
Feige 65						&	53.5*	 	& 	2		&	$c$	&	4.1		&	1.3	&	--11.9	&	1.3	&	PPMXL	&	...		&	0.35		&	0.09\\%yes
HD 205805					&	--71.0*	 	& 	4	&	$c$	&	76.7		&	1.1	&	--10.2	&	0.8	&	PPMXL	&	...		&	0.15		&	0.04\\%yes
HD 4539						&	--4*			& 	2	&	$c$	&	5.09		&	1.50&	25.19	&	1.00&	PPMXL	&	...		&	0.15		&	0.04\\%yes
HE 0004--2737					&	31.56\S	& 	28.91&	$b$	&	17.5		&	3.0	&	--11.5	&	2.8	&	NOMAD	&	4.8$c$ 	&	0.62		&	 0.09\\%
HE 0151--3919					&	--48*			& 	15	&	$e$	&	--9.2	&	9.4	&	--41.4	&	9.4	&	PPMXL	&	...		&	1.07		&	0.27\\%yes
HE 0407--1956					&	59*			& 	30	&	$d$	&	8.5		&	4.1	&	2.1		&	4.1	&	PPMXL	&	3.8$c$ 	&	0.86		&	0.12\\%
HE 1318--2111\footnotemark[2]		&	48.9		&	0.7	&	$f$	&	2.7		&	12.3	&	--1.1	&	12.3	&	NOMAD	&	3.3$c$ 	&	2.04		&	0.28\\%
HE 2135--3749\footnotemark[2]		&	45.0		&	0.5	&	$i$	&	17.1		&	9.8	&	--1.2	&	9.8	&	PPMXL	&	4.7$c$ 	&	0.66		&	0.10\\%
HE 2337--2944					&	7		& 	10	&	$d$	&	19.9		&	4.3	&	--7.3	&	4.3	&	PPMXL	&	4.5$c$ 	&	0.90		&	0.13\\%
HS 0232+3155					&	--11*			& 	30	&	$c$	&	2.8		&	4.3	&	--1.9	&	4.3	&	PPMXL	&	4.1$c$ 	&	1.95		&	0.27\\%y
HS 0233+3037					&	--129	*	& 	30		&	$c$	&	8.2		&	4.3	&	--11.4	&	4.3	&	PPMXL	&	4.3$c$ 	&	1.23		&	0.17\\%y
HS 0546+8009					&	7*			& 	30	&	$c$	&	6.5		&	4.1	&	1.2		&	4.1	&	PPMXL	&	4.1$c$ 	&	1.10		&	0.15\\%y
HS 0815+4243					&	41*			& 	30	&	$c$	&	4.2		&	5.4	&	--5.2	&	5.4	&	PPMXL	&	4.6$c$ 	&	2.60		&	0.35\\%y
HS 1236+4754					&	--46.6	& 	1.1	&	$g$	&	--14		&	1	&	2		&	6	&	NOMAD	&	4.1$c$ 	&	1.99		&	0.27\\%y
HS 1320+2622					&	--110	*	& 	30	&		$c$	&	--8.7	&	4.7	&	--1.2	&	4.7	&	PPMXL	&	4.1$c$ 	&	3.14		&	0.43\\%y
HS 1739+5244					&	--28*			&	30	&	$c$	&	--2		&	2	&	--6		&	1	&	NOMAD	&	3.5$c$ 	&	1.76		&	0.24\\%y
HS 1741+2133					&	--84*			& 	30	&	$c$	&	--13.2	&	5.1	&	--5.2	&	5.1	&	PPMXL	&	3.1$c$ 	&	1.95		&	0.27\\%y
HS 2156+2215					&	--22*			& 	30	&	$c$	&	--2.5	&	4.6	&	--2.3	&	4.6	&	NOMAD	&	3.0$c$ 	&	3.05		&	0.42\\%y
HS 2201+2610					&	--31*			& 	30	&	$c$	&	--0.7	&	4.1	&	--6.1	&	4.1	&	PPMXL	&	4.0$c$ 	&	0.95		&	0.13\\%y
HS 2208+2718					&	124*			& 	30	&	$c$	&	4.3		&	4.8	&	--7.2	&	4.8	&	PPMXL	&	3.9$c$ 	&	1.28		&	0.18\\%y
HS 2242+3206					&	--168*		& 	30	&		$c$	&	28.2		&	3.3	&	6.9		&	3.3	&	PPMXL&	4.3$c$ 	&	1.48		&	0.20\\%y
KPD 0054+5406	&	--35.5		& 	5.0	&				$h$	&	--2.0	&	0.4	&	--0.4	&	0.4	&	PPMXL&	...		&	0.57		&	0.14\\%
KPD 2040+3955\footnotemark[2]	&	--16.4	& 	1.0	&	$b$	&	--8.9	&	5.6	&	--10.2	&	5.6	&	NOMAD	&	...		&	1.03		&	0.26\\%y
KPD 2215+5037\footnotemark[2]	&	--7.2	& 	1.0	&		$b$	&	16		&	12	&	18		&	4	&	NOMAD	&	...		&	0.48		&	0.12\\%y
KUV 16256+4034\footnotemark[2]	&	--90.9	& 	0.9	&	$f$	&	--19.3	&	0.8	&	--13.2	&	0.6	&	NOMAD	&	...		&	0.48		&	0.12\\%y
PG 0004+133		&	--1.88	& 	3.9	&	$b$	&	--0.5	&	5.0	&	--20.7	&	5.2	&	NOMAD	&	...		&	0.47		&	0.12\\%y
PG 0005+179		&	--15.99\S	& 	32.94	&	$b$	&	19.2		&	5.0	&	--2.0	&	5.3	&	NOMAD	&	...		&	0.83		&	0.21\\%y
PG 0919+273\footnotemark[2]		&	--68.6	& 	0.6	&	$b$	&	23.1		&	0.7	&	--25.5	&	1.2	&	NOMAD	&	...		&	0.44		&	0.11\\%y
PG 0934+186\footnotemark[2]		&	7.7		& 	3.2	&	$b$	&	--20.7	&	4.4	&	--6.2	&	4.7	&	NOMAD	&	...		&	0.55		&	0.14\\%y
PG 1136-003\footnotemark[2]		&	63		&	14	& 	$f$&	--6.4	&	5.5	&	--20.7	&	5.5	&	NOMAD	&	...		&	0.87		&	0.23\\%yes
PG 1230+052\footnotemark[2]		&	43.1		&	 0.7	&	$b$	&	--3.2	&	5.8	&	--20.2	&	6.4	&	NOMAD	&	...		&	0.69		&	0.17\\%y
PG 1244+113\footnotemark[2]		&	7.4		&	 0.8	&				$b$	&	0.0		&	4.9	&	--8.5	&	5.3	&	NOMAD	&	...		&	0.34		&	0.09\\%y
PG 1403+316\footnotemark[2]		&	--2.1	&	 0.9	&		$b$	&	--34.3	&	2.1	&	4.5		&	2.1	&	NOMAD	&	...		&	0.98		&	0.25\\%y
PG 1519+640\footnotemark[2]		&	0.9		&	 0.8	&	$b$	&	28.1		&	2.3	&	41.2		&	2.5	&	NOMAD	&	...		&	0.39		&	0.10\\%y
PG 1558-007\footnotemark[2]		&	--71.9	&	 0.7	&	$b$	&	--4.4	&	5.6	&	--7.9	&	5.5	&	NOMAD	&	...		&	0.99		&	0.25\\%y
PG 1648+536\footnotemark[2]		&	--69.9	&	 0.9	&	$f$	&	--3.8	&	4.2	&	0.1		&	4.2	&	PPMXL	&	...		&	1.19		&	0.30\\%y
PG 2331+038\footnotemark[2]		&	--9.5	&	 1.1	&		$b$	&	--9.2	&	4.6	&	6.5		&	4.7	&	PPMXL	&	...		&	0.84		&	0.21\\%
PHL 932						&	18		&	 2	&	$c$	&	36.1		&	2.9	&	7		&	2	&	NOMAD	&	...		&	0.35		&	0.09\\%yes

\hline
\end{tabular}\\
{$a.$ \cite{Gontcharov06}, $b.$ \cite{copper11}, $c.$ \cite{edelmannphd}, $d.$ \cite{beers92}, $e.$ \cite{beers01}, $f.$ \cite{muchfuss15}, $g.$ \cite{sper02}, $h.$ \cite{downes86}, $i.$ \cite{karl06}}
\footnote[0]{\S These stars sow signs of orbital variation, the the data is insufficient to extract $\gamma$. Here the mean radial velocities and standard deviations are used.}	
\label{A3}
\end{minipage}
\end{table*}

%---------------------------------------------------------------------------------------------------------------------------------------
\begin{landscape}
\begin{table}
%\begin{minipage}{180mm}
\caption[]{Orbital parameters, galactic velocities {and 1--$\sigma$ errors }for the intermediate helium--rich stars. The last column shows the Galactic population in which each star has been classified. TH = thin disk, TK = thick disk and H = halo. }
\centering
\vspace*{0.3cm}

\begin{tabular}{p{2.3cm}cccccccccccccccccccc}
\hline

Star	&	$R_{\rm a}$&	$\pm$&	$R_{\rm p}$&	$\pm$&	$z_{\rm max}$&	$\pm$&	$e$&		$\pm$&	$J_z$&	$\pm$&$z_{\rm n}$&$\pm$&$U$ &$\pm$&$V$ &$\pm$&$W$ &$\pm$& pop\\
	& {kpc}		& 	& kpc		& & kpc		& 	&	&				&	{kpc \kmsec}&		&&&\kmsec	&		&\kmsec&			&\kmsec&		&\\
\hline
UVO 0512--08&9.22&0.20&8.03&0.12&0.54&0.12&0.07&0.01&2059.50&23.50&0.07&0.01&--23.9&3.82&248.53&18.42&--31.29&3.36&TH \\
BPS CS 22946--0005&8.93&0.59&7.42&1.53&2.04&0.29&0.09&0.10&1853.26&219.21&0.27&0.07&--13.06&23.90&224.38&20.36&63.67&3.61&TK\\ 
BPS CS 22956--0094&9.11 &	0.28	&3.85	&0.57&	0.43&	0.08&	0.41&	0.07&	1361.16&	128.91&	0.05	&0.01	&  87.27	&	11.08&		178.99&		9.73&	-4.03 &     3.07&TK \\
CPD--20$^{\circ}$1123&9.47&0.09&7.48&0.26&0.16&0.04&0.12&0.02&2024.59&31.41&0.02&0.00&--36.57&4.64&242.33&17.12&9.83&2.22&TH\\ 
HD 127493&8.36&0.04&7.14&0.22&0.18&0.02&0.08&0.01&1878.41&30.77&0.02&0.00&--7.92&2.76&234.80&18.07&11.27&3.85&TH \\
HE 1135--1134&9.38&2.19&6.67&2.14&1.75&0.52&0.17&0.17&1820.26&376.82&0.20&0.10&43.25&47.84&226.41&27.68&12.20&9.35&TK\\ 
HE 1136--2504&8.42&0.11&4.50&0.46&0.60&0.09&0.30&0.05&1450.76&89.45&0.09&0.01&--30.86&11.37&183.83&18.69&13.59&4.94&TH \\
HE 1238--1745&16.25&4.18&7.18&0.46&2.08&0.75&0.39&0.12&2353.45&192.97&0.17&0.09&--99.14&32.01&301.97&17.78&19.34&7.72&TK \\
HE 1256--2738&7.62&0.65&0.59&0.17&3.11&0.41&0.86&0.04&--112.76&470.87&0.58&0.20&19.56&68.96&--4.39&21.24&--37.87&5.24&H \\
HE 1310--2733&7.84&0.20&5.30&1.14&1.14&0.37&0.19&0.10&1516.45&172.76&0.20&0.07&--19.49&25.31&204.22&23.06&37.22&3.34&TH \\
HE 2218--2026&19.52&4.69&1.30&0.66&18.34&5.68&0.87&0.08&194.61&201.52&1.91&2.33&303.88&39.04&38.31&13.62&96.10&13.19&H \\
HE 2357--3940&8.71&0.23&8.32&0.05&0.34&0.22&0.02&0.02&2049.08&14.24&0.04&0.03&4.24&3.79&253.41&20.50&22.22&8.09&TH \\
HE 2359--2844&8.38&0.05&1.92&1.10&4.39&0.58&0.63&0.17&604.14&328.53&1.94&5.16&--30.98&26.23&115.55&14.53&93.06&5.27&TK\\ 
HS 1051+2933&19.33&3.26&3.33&0.86&3.76&0.54&0.71&0.08&1437.27&284.00&0.20&0.05&--227.82&22.18&163.53&7.65&--55.98&2.23&H\\ 
JL 87&9.59&	0.42	&8.01	&0.15	&0.70	&0.13&	0.09	&0.03&	2084.43&	35.84	&0.08&	0.02&28.63&10.31&257.23&20.95&3.86&6.08&TH \\
\lsfour&8.12&0.20&1.19&2.28&0.26&0.38&0.74&0.33&--523.81&476.48&0.03&0.05&15.84&31.97&--53.99&29.73&--7.89&27.66&H\\ 
PG 0229+064&10.69&0.41&8.40&0.06&0.31&0.07&0.12&0.02&2248.48&36.62&0.04&0.01&--23.61&3.10&270.10&19.24&--7.01&3.13&TH\\ 
PG 0240+046&10.18&0.28&4.36&0.62&0.61&0.09&0.40&0.07&1524.02&134.50&0.06&0.01&88.99&16.76&181.23&12.97&--8.31&3.28&TK \\
PG 0909+276&9.36&0.35&9.00&0.10&0.80&0.12&0.02&0.02&2172.43&40.45&0.09&0.01&--4.36&6.63&249.77&17.57&22.41&2.30&TH \\
SB 705&10.52&	0.98	&7.87&	0.26	&0.39	&0.07&	0.14	&0.06&	2161.18	&41.51&	0.05&	0.05&--22.65&5.64&265.48&19.55&6.53&2.34&TH \\
TON 107&9.28&0.69&3.54&1.92&3.32&1.86&0.45&0.21&1125.47&481.73&0.36&0.20&--47.73&40.41&139.93&11.34&72.51&2.43&TK\\ 
$\rm UVO\ 0825+15$&14.93&0.58&8.44&0.01&0.86&0.06&0.28&0.02&2543.88&35.38&0.09&0.01&--19.81&2.30&307.23&19.44&--31.7&2.46&	TK \\
SDSS J092440.11 +305013.16&9.22&0.19&6.11&1.27&0.75&0.12&0.20&0.10&1783.13&199.25&0.09&0.02&--20.87&17.95&205.01&17.50&4.70&2.31&TH \\
SDSS J160131.30 +044027.00&8.10&0.26&5.21&0.81&0.81&0.21&0.22&0.07&1546.31&154.44&0.10&0.03&38.65&12.31&208.68&24.01&30.77&2.90&TH \\
SDSS J175137.44 +371952.37&8.18&0.27&3.56&0.47&0.76&0.32&0.39&0.05&1229.67&127.13&0.10&0.04&43.05&17.60&163.55&25.19&23.15&4.71&TK \\
SDSS J175548.50 +501210.77&8.91&0.18&4.83&0.17&0.24&0.03&0.30&0.02&1561.31&28.15&0.03&0.00&58.29&3.29&196.40&22.87&--0.59&3.23&TH \\
HS 1000+471&15.24	&9.04&	10.75	&2.11&	5.01	&6.46	&0.17	&0.20	&2672.06	&956.10	&0.37&	0.58&13.11&85.32&138.81&12.14&7.36&11.62&TK \\

\hline
\end{tabular}
\label{heinter}
%\end{minipage}
\end{table}
\end{landscape}

\begin{landscape}
\begin{table}
\caption[]{Orbital parameters and galactic velocities for the extreme helium--rich stars.}
\centering
\vspace*{0.3cm}

\begin{tabular}{p{2.3cm}cccccccccccccccccccc}
\hline

Star	&	$R_{\rm a}$&	$\pm$&	$R_{\rm p}$&	$\pm$&	$z_{\rm max}$&	$\pm$&	$e$&		$\pm$&	$J_z$&	$\pm$&$z_{\rm n}$&$\pm$&$U$ &$\pm$&$V$ &$\pm$&$W$ &$\pm$& pop\\
	& kpc		& 	& kpc		& & kpc		& 	&	&				&	kpc \kmsec&		&&&\kmsec	&		&\kmsec&			&\kmsec&		&\\
\hline
BPS CS 22940--0009&7.90	&0.15&	4.68&	0.85&	0.53&	0.19&	0.26	&0.08&	1452.08&	160.86&	0.08	&0.03	&-21.95	&17.29&	194.19&	16.63&	-17.91  &    4.52&TH \\
BPS CS 29496--0010&14.87&	3.14&	7.17&	0.46&	3.03	&1.01	&0.35&	0.09&	2233.39&	184.77	&0.21	&0.09	& -84.03&	12.11&	282.43&	26.78&	75.28  &   2.78&H \\
HE 0001--2443&8.46	&0.06&	4.16	&0.58	&0.65	&0.10&	0.34	&0.06	&1383.07	&125.65	&0.10&	0.02&--22.14&4.40&210.48&14.34&--1.11&3.63&TH \\
HE 0342--1702&10.26&0.93&7.48&0.85&0.59&0.10&0.16&0.05&2077.65&165.70&0.08&0.01&--50.94&16.11&243.34&21.08&4.94&4.29&TH \\
HE 1251+0159&11.10&0.92&3.72&0.72&2.09&0.62&0.50&0.09&1350.89&197.91&0.22&0.09&--137.71&15.89&175.63&17.17&--46.38&11.58&TK\\ 
LB 3229&8.38	&0.07	&5.51&	0.99	&0.57	&0.10&	0.21	&0.09	&1630.99&	168.18	&0.07	&0.01	& 9.08&18.79	&204.59&	23.30&	1.14   &   12.97&TH \\
PG 0039+135&8.97&0.24&6.10&1.36&1.89&0.95&0.19&0.12&1687.63&321.38&0.29&0.15&--37.5&9.78&206.38&35.29&78.71&29.10&TK\\ 
PG 1413+114&8.80&0.90&2.33&1.17&2.03&0.36&0.58&0.16&873.22&368.26&0.26&0.07&--103.25&31.16&125.49&16.15&--12.69&7.93&TK\\ 
PG 1536+690&11.54&0.71&0.99&1.04&9.76&1.48&0.84&0.15&222.93&108.13&1.29&1.05&--141.57&13.37&33.63&25.86&--148&4.64&H \\
PG 2321+214&8.74&0.19&5.59&0.46&0.35&0.14&0.22&0.04&1680.81&75.47&0.04&0.02&34.74&12.73&207.73&14.91&--14.52&4.34&TH \\
PG 2352+181&8.59&0.15&5.10&1.22&0.49&0.42&0.26&0.11&1579.87&236.22&0.06&0.05&25.36&12.19&195.09&24.67&25.45&14.90&TH \\
PG 0902+057&9.25&0.99&8.96&0.41&1.61&0.60&0.02&0.04&2103.74&147.58&0.18&0.07&--3.01&12.49&243.27&22.74&--62.28&2.23&TK \\
PG1615+413&9.03&1.61&1.69&1.57&3.18&0.53&0.68&0.27&633.66&445.89&0.44&0.16&--100.82&47.61&87.85&19.11&18.77&4.74&TK \\
PG1600+171&7.25&0.81&1.36&1.15&2.39&0.51&0.68&0.23&386.13&484.78&0.36&0.10&61.00&51.03&68.97&27.47&32.72&8.36&TK \\
PG1658+273&8.35&1.02&2.63&0.33&5.76&2.44&0.52&0.03&632.82&310.84&1.24&3.28&60.61&28.90&95.49&34.81&150.84&4.03&TK \\
PG1715+273&7.66&0.57&6.30&0.85&1.78&0.33&0.10&0.05&1593.18&185.26&0.27&0.06&--34.72&22.37&227.38&5.85&30.17&7.88&TK\\ 
HS1844+637&11.36&1.70&7.40&2.29&2.61&2.36&0.21&0.20&2037.64&342.34&0.25&0.25&60.76&59.10&236.59&25.46&74.93&10.65&TK \\
PG1554+408&10.03&2.88&7.89&0.90&3.77&0.81&0.12&0.16&1867.92&380.46&0.47&0.25&3.93&49.11&243.61&22.08&101.52&6.73&TK \\
PG2258+155&9.76&1.77&8.40&0.60&1.62&0.47&0.07&0.09&2094.75&228.02&0.17&0.05&7.70&35.24&256.79&17.59&--44.41&6.27&TK \\
PG1127+019&8.51&0.08&6.22&0.60&0.56&0.03&0.15&0.05&1751.48&87.32&0.07&0.01&--12.56&10.43&215.77&18.88&5.11&2.29&TH \\
PG1415+492&10.56&1.65&8.48&0.06&2.24&0.20&0.11&0.08&2142.10&167.49&0.21&0.04&--3.01&20.76&261.89&18.19&64.48&4.48&TK \\
PG2215+151&12.30&2.04&7.51&0.50&1.58&0.79&0.24&0.09&2192.41&117.95&0.13&0.07&--233.78&12.14&215.86&20.93&31.20&16.51&TK \\
PG1544+488&11.23&0.80&4.34&0.45&1.05&0.32&0.44&0.07&1554.39&81.96&0.09&0.03&33.56&14.72&259.72&20.51&--24.84&6.21&TK \\

\hline
\end{tabular}
\end{table}
\end{landscape}

\begin{landscape}
\begin{table}
\caption[]{Orbital parameters and galactic velocities for the helium--deficient stars. }
\centering
\vspace*{0.3cm}

\begin{tabular}{p{2.3cm}cccccccccccccccccccc}
\hline

Star	&	$R_{\rm a}$&	$\pm$&	$R_{\rm p}$&	$\pm$&	$z_{\rm max}$&	$\pm$&	$e$&		$\pm$&	$J_z$&	$\pm$&$z_{\rm n}$&$\pm$&$U$ &$\pm$&$V$ &$\pm$&$W$ &$\pm$& pop\\
	& kpc		& 	& kpc		& & kpc		& 	&	&				&	kpc \kmsec&		&&&\kmsec	&		&\kmsec&			&\kmsec&		&\\
\hline
CD --35$^{\circ}$15910&9.87&0.26&7.66&0.13&0.29&0.05&0.13&0.02&2080.52&13.91&0.04&0.01&--32.24&2.57&265.86&18.34&--16.22&4.28&TH \\
Feige 65&9.40&0.20&8.37&0.06&1.50&0.12&0.06&0.01&2061.68&29.80&0.17&0.01&--14.71&5.48&252.57&15.06&64.24&3.82&TH \\
HD 205805&10.75&0.25&6.64&0.32&0.44&0.18&0.24&0.03&1994.73&31.93&0.05&0.02&76.20&5.68&243.01&14.31&24.66&2.13&TH \\
HD 4539&9.51&0.30&8.45&0.01&0.33&0.07&0.06&0.02&2145.20&27.70&0.04&0.01&0.85&4.73&262.19&19.06&21.12&2.22&TH \\
HE 0004--2737&8.40	&0.08	&5.54&	0.64	&0.88&	0.37	&0.20&	0.06&	1620.18&	98.79	&0.11&	0.05&18.14&11.70&201.24&21.91&--4.7&15.62&TH\\ 
HE 0407--1956&9.29&0.44&6.40&1.13&0.61&0.20&0.18&0.09&1841.28&171.99&0.07&0.02&33.66&19.40&215.41&25.04&--8.82&12.76&TH\\ 
HE 1318--2111&9.53&5.60&5.38&2.70&2.01&1.90&0.28&0.29&1609.83&688.11&0.25&0.37&--95.15&97.41&223.22&22.12&28.68&21.34&TK \\
HE 2135--3749&8.46&2.28&7.95&0.29&1.36&0.59&0.03&0.13&1920.73&244.94&0.16&0.08&--0.18&30.72&248.84&16.15&--61.42&8.68&TH \\
HE 2337--2944&8.69	&0.36	&4.76&	0.86	&0.97	&0.18	&0.29	&0.08&	1501.14&	170.00	&0.12&	0.02&46.28&21.90&217.37&18.81&--55.94&9.91&TH \\
HS 0232+3155&10.09&0.58&7.93&2.85&0.86&0.19&0.12&0.18&2112.29&379.73&0.09&0.02&18.99&37.19&221.40&25.13&7.12&9.79&TH \\
HS 0233+3037&10.08&0.53&2.83&0.82&0.88&0.76&0.56&0.10&1119.88&253.56&0.19&0.28&--79.21&22.73&129.70&22.24&33.28&14.04&TK\\ 
HS 0546+8009&9.46&1.78&8.93&0.90&1.02&0.61&0.03&0.09&2163.23&227.46&0.11&0.07&7.14&23.50&245.75&25.45&40.98&11.09&TH \\
HS 0815+4243&10.73&1.58&6.74&4.45&2.81&1.66&0.23&0.36&1858.99&704.73&0.42&1.64&7.17&58.53&184.18&20.76&63.79&6.19&TK \\
HS 1236+4754&11.11&2.20&4.98&1.76&3.61&1.83&0.38&0.09&1531.28&398.46&0.71&0.45&110.13&18.86&183.01&44.30&--49.97&14.57&TK\\ 
HS 1320+2622&9.61&1.99&5.12&2.98&5.83&1.91&0.31&0.26&1185.05&573.89&0.73&0.51&96.67&50.97&154.91&33.21&--84.9&5.15&TK \\
HS 1739+5244&8.44&0.12&6.54&1.55&0.97&0.14&0.13&0.11&1767.63&223.17&0.13&0.03&--16.44&11.67&221.73&21.14&3.42&16.42&TH \\
HS 1741+2133&7.42&0.36&1.91&1.31&1.55&0.81&0.59&0.23&712.46&296.68&0.63&1.67&26.96&30.78&106.93&33.56&60.31&10.77&TK \\
HS 2156+2215&8.58&1.18&7.26&2.71&1.37&0.57&0.08&0.21&1848.99&331.22&0.17&0.08&21.09&52.93&230.76&25.79&16.04&4.57&TH \\
HS 2201+2610&8.34&0.04&6.20&1.68&0.35&0.05&0.15&0.13&1743.14&237.88&0.04&0.01&--3.34&16.18&217.48&25.30&2.05&12.47&TH \\
HS 2208+2718&23.93&7.25&8.34&0.09&4.52&2.16&0.48&0.12&2841.56&251.09&0.19&0.11&21.62&32.43&347.42&19.72&--88.06&10.91&H \\
HS 2242+3206&11.50&1.37&0.53&0.69&0.93&0.22&0.91&0.11&234.00&270.30&0.08&0.02&179.27&26.30&37.35&19.43&18.52&3.32&TK \\
KPD 0054+5406&8.95&0.07&7.12&0.31&0.16&0.02&0.11&0.02&1930.35&44.49&0.02&0.00&--22.44&3.32&229.16&18.52&11.34&4.03&TH \\
KPD 2040+3955&9.07&0.77&6.64&0.51&0.15&0.41&0.15&0.08&1873.78&14.64&0.02&0.05&--44.22&18.04&233.62&20.29&11.49&3.21&TH \\
KPD 2215+5037&9.66&0.68&6.80&0.78&0.33&0.30&0.17&0.09&1940.89&59.13&0.04&0.03&53.44&25.75&237.36&13.11&20.98&5.95&TH \\
KUV 16256+4034&8.27&0.04&3.41&0.37&0.57&0.05&0.42&0.04&1207.94&89.08&0.08&0.01&9.62&9.55&155.33&19.50&--25.57&4.36&TK \\
PG 0004+133&8.72&0.21&6.78&0.58&0.50&0.22&0.13&0.04&1852.57&91.01&0.06&0.03&--25.13&9.38&226.44&15.36&--20.8&3.50&TH \\
PG 0005+179&9.36	&0.73	&5.23	&1.28	&0.59&	0.17&	0.28&	0.11&	1653.79&	247.74	&0.06	&0.02&61.03&28.17&201.31&16.06&--0.41&3.57&TH \\
PG 0919+273&11.27&0.17&5.60&0.56&0.48&0.01&0.34&0.05&1844.09&99.11&0.07&0.00&--104.6&4.37&220.52&10.22&--15.73&4.92&TK\\ 
PG 0934+186&9.03&0.37&7.47&0.63&0.68&0.34&0.09&0.04&1962.96&101.55&0.08&0.04&22.27&6.46&233.42&22.38&--26.91&2.05&TH \\
PG 1136--003&8.60&0.21&3.16&0.98&0.80&0.37&0.46&0.12&1150.28&242.10&0.14&0.09&--52.42&14.98&231.22&17.93&--86.17&8.78&TK\\ 
PG 1230+052&8.62&0.26&4.39&1.00&0.84&0.13&0.33&0.10&1431.81&201.26&0.14&0.03&--45.5&16.48&255.43&21.55&3.96&5.58&TH \\
PG 1244+113&8.67&0.24&7.50&0.50&0.40&0.08&0.07&0.03&1946.10&73.24&0.05&0.01&--21.4&8.78&241.18&17.36&10.60&2.46&TH \\
PG 1403+316&10.42&0.93&3.49&0.87&1.58&0.30&0.50&0.12&1284.22&216.57&0.17&0.03&122.82&6.83&166.06&33.84&47.86&3.51&TK \\
PG 1519+640&16.20&2.95&8.32&0.05&1.84&1.01&0.32&0.08&2565.16&127.65&0.13&0.08&34.15&16.78&312.14&20.38&--52.7&8.42&TK \\
PG 1558--007&7.92&0.38&5.01&1.24&1.03&0.41&0.23&0.11&1483.40&229.31&0.14&0.06&39.30&18.44&204.22&19.91&--39.82&4.73&TH \\
PG 1648+536&8.42&0.18&4.75&0.76&0.93&0.25&0.28&0.07&1482.03&129.92&0.12&0.04&22.73&19.05&187.00&22.50&--21.23&6.37&TH \\
PG 2331+038&11.91&2.12&8.37&0.12&1.32&0.59&0.17&0.09&2303.87&145.96&0.14&0.07&10.68&16.74&263.73&25.22&7.22&6.23&TH \\
PHL 932&9.80&0.34&6.79&0.55&0.27&0.09&0.18&0.05&1952.07&62.56&0.03&0.01&55.02&14.12&236.66&18.74&5.29&9.70&TH \\

\hline
\end{tabular}
\label{hedef}
\end{table}
\end{landscape}

\bsp
\label{lastpage}

\end{document}